\documentclass[12pt,graphicx,amsmath]{article}
\usepackage{amsmath}%
\usepackage{amsfonts}%
\usepackage{amssymb}%
\usepackage{eucal}
\usepackage{makeidx}
\usepackage{afterpage}
\usepackage{pdflscape}
\makeindex
\usepackage{nomencl}
\makenomenclature
\usepackage{diagbox}
\usepackage{theorem}
\usepackage{url}
\usepackage{array}

\usepackage{upgreek}
\usepackage{epsfig}
\usepackage{graphicx}
\usepackage{verbatim}
\usepackage{float}
\usepackage{braket}
\usepackage{vector}
\usepackage{appendix}
\usepackage[margin=2.5cm]{geometry}
\usepackage{parskip}

\usepackage[titles]{tocloft}
\usepackage{pst-node}
\usepackage{amsmath}
\newcommand{\qed}{\nobreak \ifvmode \relax \else
      \ifdim\lastskip<1.5em \hskip-\lastskip
      \hskip1.5em plus0em minus0.5em \fi \nobreak
      \vrule height0.75em width0.5em depth0.25em\fi}
\def\VZ#1{\phantom{-}\ifx\relax#1\relax0\else\rnode{#1}{0}\fi}
\usepackage{blkarray}
\usepackage{enumerate}

\newcommand{\cc}[2]{c{#1\atopwithdelims[]#2}}
\newcommand{\CC}[2]{C{#1\atopwithdelims[]#2}}
\newcommand{\nn}{\nonumber}
\newcommand{\ba}{\begin{eqnarray}}
\newcommand{\ea}{\end{eqnarray}}
\def\beq{\begin{equation}}
\def\eeq{\end{equation}}
\def\beqn{\begin{eqnarray}}
\def\eeqn{\end{eqnarray}}

\def\AEF{A.E. Faraggi}

\def\IJMP#1#2#3{{\it Int.\ J.\ Mod.\ Phys.}\/ {\bf A#1} (#2) #3}

\def\EJP#1#2#3{{\it Eur.\ Phys.\ Jour.}\/ {\bf C#1} (#2) #3}

\def\JHEP#1#2#3{{\it JHEP}\/ {\bf #1} (#2) #3}
\def\NPB#1#2#3{{\it Nucl.\ Phys.}\/ {\bf B#1} (#2) #3}
\def\PLB#1#2#3{{\it Phys.\ Lett.}\/ {\bf B#1} (#2) #3}
\def\PRD#1#2#3{{\it Phys.\ Rev.}\/ {\bf D#1} (#2) #3}
\def\PRL#1#2#3{{\it Phys.\ Rev.\ Lett.}\/ {\bf #1} (#2) #3}

\def\AMP#1#2#3{{\it Adv.\ Math.\ Phys.}\/ {\bf#1} (#2) #3}
\def\LNP#1#2#3{{\it Lect.\ Notes.\ Phys.}\/ {\bf #1} (#2) #3}
\def\etal{{\it et al\/}}
\def\SO64{$SO(6)\times SO(4)$}
\def\SU51{$SU(5)\times U(1)$}
\def\S321{$SU(3)\times U(1)_C\times SU(2)\times U(1)_L$}

\begin{document}
\vspace{80mm}
\vspace{40mm}

\begin{center}
\LARGE
{\bf{Classification\\ \medskip of Standard-Like Heterotic-String Vacua}}
\end{center}
\begin{center}

\end{center}
\begin{center}
\Large
Alon E. Faraggi$^{1}$\footnote{alon.faraggi@liv.ac.uk}, 
John Rizos$^2$\footnote{irizos@uoi.gr} and
Hasan Sonmez$^{1}$\footnote{Hasan.Sonmez@liv.ac.uk}
\end{center}

\vspace{10mm}
\begin{center}

$^{1}$ 
\emph{Dept. of Mathematical Sciences, University of Liverpool, 
Liverpool L69 7ZL, UK}\\
\vspace{.05in}
{\it $^{2}$ Department of Physics,
              University of Ioannina, GR45110 Ioannina, Greece\\}
\vspace{.025in}

\end{center}
\vspace{20mm}

\begin{abstract}
\normalsize
\noindent

We extend the free fermionic classification methodology to the class of 
standard--like heterotic--string vacua, in which the $SO(10)$ GUT symmetry is
broken at the string level to $SU(3)\times SU(2)\times U(1)^2$. The space
of GGSO free phase configurations in this case is vastly enlarged compared
to the corresponding $SO(6)\times SO(4)$ and $SU(5)\times U(1)$ vacua. 
Extracting substantial numbers of phenomenologically viable models
therefore requires a modification of the classification methods. 
This is achieved by identifying conditions on the GGSO projection coefficients,
which are satisfied at the $SO(10)$ level by random phase configurations, and
that lead to three generation models with the $SO(10)$ symmetry broken to 
the $SU(3)\times SU(2)\times U(1)^2$ subgroup. Around each of these fertile 
$SO(10)$ configurations, we perform a complete classification of 
standard--like models, by adding the $SO(10)$ symmetry breaking basis 
vectors, and scanning all the associated GGSO phases. Following this 
methodology we are able to generate some $10^7$ three generation Standard--like 
Models. We present the results of the classification and one exemplary
model with distinct phenomenological properties, compared to previous 
SLM constructions.  

\end{abstract}

\newpage
\tableofcontents
\normalsize
\bigskip

\section{\emph{Introduction}}
The Standard Model utilises the 
framework of perturbative quantum field theories and 
provides viable perturbative 
parameterisation of 
all subatomic observational data up to the electroweak 
symmetry breaking scale. The synthesis of gravity with the 
gauge interactions requires, however, a departure from 
perturbative quantum field theories. String theories 
provide a consistent approach to perturbative quantum 
gravity. Furthermore, the consistency conditions 
espouse gravity with the gauge and matter components of the 
subatomic world. By doing that string 
theory provides the ingredients for the 
development of a phenomenological approach to 
quantum gravity. While this approach is still in 
its infancy, the development of deeper understanding 
of the theory, as well as of the tools for the analysis 
of phenomenological vacua is required. 

Indeed, since the early realisation that string theory provides 
the ingredients for the unification of gravity and the 
gauge interactions \cite{candelas, heterotic}, a 
variety of target--space and worldsheet tools 
have been utilised to construct phenomenological 
string vacua \cite{spreview}. The models 
constructed in the free fermionic formulation \cite{fff}
correspond to toroidal $Z_2\times Z_2$ orbifold compactifications
at special points in the moduli space with discrete 
Wilson lines \cite{z2xz2}.
The early constructions in the late eighties 
were obtained by a ``trial and error''
method and corresponded to asymmetric $Z_2\times Z_2$ 
orbifold compactifications, in which the 
observable $E_8$ gauge symmetry is 
broken to an $SO(10)$ subgroup. 
Quasi--realistic three generation models with 
$SU(5)\times U(1)$ (flipped $SU(5)$) \cite{fsu5}, 
$SU(3)\times SU(2)\times U(1)^2$ (Standard--like) \cite{slm}, 
$SO(6)\times SO(4)$ (Pati--Salam) \cite{alr}, 
and 
$SU(3)\times SU(2)^2\times U(1)$ (left--right symmetric) \cite{lrs}
$SO(10)$ subgroups gave rise to quasi--realistic examples, whereas the 
case with $SU(4)\times SU(2)_L\times U(1)_R$
was shown not to produce realistic models \cite{su421}. 
The early free fermionic models
shared an underlying NAHE--based structure \cite{nahe} 
and consisted of a small number of examples. 
More recent methodology pursued 
the systematic classification of large spaces 
of free fermionic string vacua 
\cite{gkr, fknr, fkr, acfkr, su62,frs,hasan}, as well as in other 
approaches \cite{statistical}.

The free fermionic classification method was initially developed 
for type II superstring vacua in ref. \cite{gkr}.
It was extended for the classification of symmetric
$Z_2\times Z_2$ heterotic--string orbifolds with 
an unbroken $SO(10)$ gauge group in refs. \cite{fknr,fkr}. 
The classification of vacua with
$SO(6)\times SO(4)$ Pati--Salam (PS) $SO(10)$
subgroup was developed in ref. \cite{acfkr}, 
and the case with the $SU(5)\times U(1)$ (FSU5)
$SO(10)$ subgroup was pursued in ref. \cite{frs, hasan}.
The classification program led to several important results.
The case with an unbroken $SO(10)$ subgroup revealed the 
existence of a new duality symmetry in the space
of heterotic--string vacua with $(2,0)$ worldsheet 
supersymmetry, akin to mirror symmetry \cite{mirror}, 
under the exchange of spinorial plus anti--spinorial 
and vectorial representations of $SO(10)$ 
\cite{fkr, cfkr}.
It was extended to compactifications corresponding to 
interacting worldsheet CFTs in ref. \cite{panos}. 
The PS classification produced examples of exophobic
heterotic--string vacua, in which
exotic states with fractional electric charge
do not appear as massless states in the physical spectrum 
\cite{acfkr}. The classification methodology 
provides an insight into the global symmetries that 
underlie the large space of vacua, as, for example, in the 
case of spinor--vector duality, as well as providing a
trawling algorithm to extract string models 
with desired phenomenological properties.
The spinor--vector duality may be a reflection of a much
wider symmetry structure that underlie the fermionic 
$Z_2\times Z_2$ orbifolds \cite{moonshine}.
Another example is the observation that 
a large space of FSU5 vacua do not contain exophobic 
models with an odd number of chiral generations \cite{frs, hasan}. 
The fishing procedure was employed to construct exophobic
three generation models with $SU(6)\times SU(2)$ Grand 
Unified Theory (GUT) \cite{su62}, as well as 
string vacua that allow for the existence of a 
light family universal $Z^\prime$ \cite{frzprime}. A general 
signature of this class of low scale $Z^\prime$ models is via
di--photon excess \cite{diphoton}.

In this paper we extend the classification methodology 
of free fermionic heterotic--string models to the 
case in which the $SO(10)$ symmetry 
is broken to the $SU(3)\times SU(2)\times U(1)^2$ 
standard--like model (SLM) subgroup.
This class of vacua introduces several novel features. 
The first is that the set of basis vectors that are 
used to span the space of models utilises both the 
PS and FSU5 symmetry breaking patterns in two 
separate basis vectors. This makes the analysis 
of the spectrum and the development of automated 
techniques to extract the physical states far more 
cumbersome compared to the previous two cases. 
The second complexity is with respect to the 
type of exotic states that arise in the spectrum 
of the standard--like models \cite{fc}. The PS as well
as the FSU5 models produce exotic states that 
carry fractional charge $\pm1/2$ and must 
therefore be sufficiently rare and/or sufficiently 
heavy \cite{vhalyo}. As the SLM models contain 
both the PS and FSU5 breaking patterns, they admit 
sectors that possess the PS or FSU5 symmetry
and therefore also give rise to states with
fractional electric charge $\pm1/2$. 
However, the SLM class of models also contain
sectors that arise from combinations 
of the PS and FSU5 breaking basis vectors \cite{slm}.
These sectors produce states that carry the standard 
charges with respect to the Standard Model subgroup,
but carry fractional charge with respect to 
the $U(1)_{Z^\prime}$, which is embedded in $SO(10)$
and is orthogonal to the Standard Model gauge group
\cite{fc}. Such states are therefore particular to the SLM--models. 
They can produce viable dark matter candidates \cite{ccr}
as well as play a role in the symmetry breaking of 
the $U(1)_{Z^\prime}$ \cite{slm} and in the stringy
see--saw mechanism \cite{seasaw}.

The inclusion of two basis vectors that break the $SO(10)$ symmetry 
increases the complexity of the classification method. In the first
instance we find that the space of a priori distinct vacua is 
increased to the order of $2^{72}$ independent configurations 
as compared to $2^{51}$ in the cases of the FSU5 and PS 
free fermionic heterotic--string vacua. To explore the space
of phenomenologically viable models this necessitates adaptation
of the classification methodology, in a two stage process. 
The first stage is a pre--selection of configurations with 
unbroken $SO(10)$ symmetry with net number of twelve generations
or larger.
Moreover, it turns out that one can constrain the space of 
pre--selected configurations with $SO(10)$ gauge symmetry that can 
lead to three generation models with $SU(3)\times SU(2)\times U(1)^2$ 
symmetry. We therefore impose these constrains on the random 
generation of free phases configurations and only scan the 
models around these fertile $SO(10)$ cores. The reason being that the 
frequency of phenomenologically viable models among the total number of models
is too small to generate a significant statistical sampling of 
phenomenologically interesting models.
Around these pre--selected fertile configurations with $SO(10)$ symmetry 
we perform a complete classification of the standard--like models
by adding the two $SO(10)$ breaking vectors and varying 
all the phases associated with the added basis vectors.
This method ensures that the number of viable three 
generation models is not too diluted and is accessible to 
the statistical sampling. This two stage process represents a 
departure from the methodology used in the classification of the 
PS and FSU5 free fermionic models. 
Additionally, and differently 
from the previous cases of the FSU5 and PS models we do not 
restrict a priori our scan to vacua in which only untwisted 
spacetime vector bosons remain in the physical spectrum. 
Therefore, the gauge symmetry may be enhanced in some of the 
models. The requirement that the observable gauge symmetry
is that of the Standard Model times some $U(1)$ symmetries
is imposed as a test on fished out models. 
The reason for this change is that the number of sectors
that produce additional vector bosons is large and 
imposing that all of those are projected out imposes
a large number of constraints and is unnecessarily cumbersome. 
We find that about 20\% of the total number of models 
contain enhanced symmetries. Viable models allow for 
enhancement of the hidden sector rank eight gauge symmetry, 
whereas enhancements of the observable sector, or mixed 
enhancements are not allowed. 

Our paper is organised as follows: in section \ref{analysis}
we introduce the free fermionic classification methodology. In section
\ref{analysisspec} we discuss the sectors that produce massless physical 
states 
in the free fermionic standard--like models. We first present the symmetry
enhancing sectors and next elaborate on the twisted sectors
that produce massless matter states. These sectors are divided 
into sectors that produce standard model observable sector states 
versus those that produce hidden sector as well as exotic states. 
Our focus in this paper is on extracting phenomenologically viable 
vacua and we discuss the special procedure adopted here to 
obtain these models. We present all the matter producing sectors
that arise in the models, but our systematic classification in
this paper is solely with respect to the observable Standard Model
states. In section \ref{tmy} we impose the existence of a leading top
quark Yukawa coupling and discuss the implementation of this 
requirement in the classification procedure. in section \ref{resu} we discuss
the outcome of our computerised search, which results in some 
$10^{7}$ three generation Standard--like Models. 
In section \ref{example} we 
present an exemplary model with several distinct 
properties as compared to the earlier SLM constructions \cite{slm}.
This demonstrates 
the power of our computerised methodology in extracting models with
specific phenomenological properties. Section \ref{conc} concludes our paper. 

\section{\emph{Standard-Like Free Fermionic Models}}
\label{analysis}

In this paper 
we extend the free fermionic classification method of 
\cite{fknr,fkr,acfkr,frs} 
to the case of vacua with the 
standard--like subgroup of $SO(10)$. 
The free fermionic model building rules are 
formulated in terms of a set of basis vectors 
and the Generalised Gliozzi--Scherk--Olive (GGSO) 
projection coefficients of the one--loop partition 
function \cite{fff}.
It facilitates straightforward analysis of the 
physical massless states and of the renormalisable and 
non--renormalisable terms in the superpotential. 
The $SO(10)$ GUT symmetry is broken directly at the string level. 
In this paper the manifest unbroken $SO(10)$ subgroup
in the low energy effective field theory is 
$SU(3)\times SU(2)\times U(1)^2$. 
The matter states that give rise to the Standard Model 
fermionic representations are obtained from spinorial $16$ 
representations of $SO(10)$ decomposed under the 
unbroken $SO(10)$ subgroup. Similarly, the light Standard Model
Higgs states arise from vectorial 10 representations of 
$SO(10)$. The free fermionic models correspond to 
$Z_2\times Z_2$ orbifold compactifications with 
${\cal N}=(2,0)$ worldsheet supersymmetry and discrete 
Wilson lines. All the models that we classify preserve the $SO(10)$ 
embedding of the weak hypercharge and possess $N=1$ 
spacetime supersymmetry. Extension to nonsupersymmetric 
vacua \cite{nonsusy} can similarly be pursued and is 
left for future work.

\subsection{\emph{The Free Fermionic Formulation}}\label{tfff}

We recap the salient features of the free fermionic construction,
essential for the ensuing discussion. Further details of the notation and
construction can be found in the literature \cite{fff, z2xz2, fsu5, slm,alr,
lrs,su421, nahe, gkr, fknr, fkr, acfkr, su62, frs, hasan}. 
In the free fermionic formulation all the extra degrees of freedom
required to generate a consistent string theory are represented in terms 
of free fermions propagating on the two dimensional string worldsheet.
In the four dimensional heterotic--string in the light--cone gauge
these include 20 left--moving and 44 right--moving real worldsheet fermions.
When parallel transported around the non--contractible 
loops of the vacuum to vacuum amplitude the worldsheet fermions 
can pick up a nontrivial phase. 
The transformation properties of the 64 worldsheet fermions are
encoded in 64 dimensional boundary condition vectors, 
$$v_i=\left\{\alpha_i(f_1),\dots,\alpha_i(f_{20})|\alpha_i(\overline{f}_1),
\dots,\alpha_i({\overline{f}_{44}})\right\}.$$ 
A string vacuum in the free fermionic formulation is specified
in terms of a set of basis vectors, $v_{1},\dots,v_{N}$, 
that must be
consistent with modular invariance constraints.
The basis vectors span a space $\Xi$ of $2^{N+1}$ sectors, 
obtained as linear combination of the basis vectors, 
\begin{equation}
\xi = \sum_{i=1}^N m_j v_i, \,\,\,\,\,\,\,\,\,\, m_j = 0,1,\dots,N_j-1,
\label{Xi}
\end{equation}
where $N_j \cdot v_j = 0$ mod $2$, and produce the 
string spectrum. The physical string states  $|S_{\xi}>$ 
in a given sector
are constrained by modular invariance, which is encoded 
in terms of the boundary condition basis vectors and 
the one--loop Generalised GSO projection (GGSO) coefficients as
\begin{equation}\label{gso}
e^{i\pi v_i\cdot F_{\xi}} |S_{\xi}> = \delta_{{\xi}}\ 
\CC{\xi}{v_i}^* 
|S_{\xi}>,
\end{equation}
where $F_{\xi}$ is the fermion number operator, and
$\delta_{{\xi}}=\pm1$ is the space--time spin statistics index.
Different choices of GGSO projection 
coefficients $\cc{{\xi}}{v_i}=\pm1; \pm i$,
consistent with modular invariance produce different models.
In summary: a model is specified by a set of 
boundary condition basis vectors 
$v_{1},\dots,v_{N}$ and a set of $2^{N(N-1)/2}$ of independent
GGSO projection coefficients $\CC{v_i}{v_j}, i>j$.

\subsection{\emph{SO(10) Models}}

The first stage in the classification entails preselecting
a string vacuum with a total net number of generations 
exceeding 12 generations. 
In the usual notation the worldsheet fermions are denoted by: 
$\psi^\mu, \chi^{1,\dots,6},y^{1,\dots,6}, \omega^{1,\dots,6}$
(left-movers) and
$\overline{y}^{1,\dots,6},\overline{\omega}^{1,\dots,6}$,
$\overline{\psi}^{1,\dots,5}$, $\overline{\eta}^{1,2,3}$, $\overline{\phi}^{1,\dots,8}$
(right-movers). Here 32 of the real right--moving fermions are 
paired into 16 complex fermions. Of those the first five complex 
fermions, denoted by $\overline{\psi}^{1,\dots,5}$, 
produce the Cartan sub--algebra
of an $SO(10)$ GUT group; the next three, denoted by 
$\overline{\eta}^{1,2,3}$, produce three $U(1)$ generators; 
and the last eight, denoted by $\overline{\phi}^{1,\dots,8}$,
produce the Cartan generators of the hidden sector gauge group.

The Standard--like Models (SLMs) in the free fermionic construction are 
produced by a set of 14 basis vectors.
The first 12 basis vectors consist of the same basis vectors
that are used in the classification of the $SO(10)$ vacua \cite{fkr}. 
These basis vectors preserve the $SO(10)$ symmetry and 
are given by
\begin{eqnarray}
v_1={\bf1}&=&\{\psi^\mu,\
\chi^{1,\dots,6},y^{1,\dots,6}, \omega^{1,\dots,6}| \nonumber\\
& & ~~~\overline{y}^{1,\dots,6},\overline{\omega}^{1,\dots,6},
\overline{\eta}^{1,2,3},
\overline{\psi}^{1,\dots,5},\overline{\phi}^{1,\dots,8}\},\nonumber\\
v_2=S&=&\{{\psi^\mu},\chi^{1,\dots,6}\},\nonumber\\
v_{2+i}={e_i}&=&\{y^{i},\omega^{i}|\overline{y}^i,\overline{\omega}^i\}, \
i=1,\dots,6,\nonumber\\
v_{9}={b_1}&=&\{\chi^{34},\chi^{56},y^{34},y^{56}|\overline{y}^{34},
\overline{y}^{56},\overline{\eta}^1,\overline{\psi}^{1,\dots,5}\},\label{basis}\\
v_{10}={b_2}&=&\{\chi^{12},\chi^{56},y^{12},y^{56}|\overline{y}^{12},
\overline{y}^{56},\overline{\eta}^2,\overline{\psi}^{1,\dots,5}\},\nonumber\\
v_{11}=z_1&=&\{\overline{\phi}^{1,\dots,4}\},\nonumber\\
v_{12}=z_2&=&\{\overline{\phi}^{5,\dots,8}\}.
\nonumber
\end{eqnarray}

The additional two basis vectors break the gauge symmetry to 
the Pati--Salam (PS) and flipped $SU(5)$ (FSU5) subgroups. 
The classification of the PS models was done in ref. 
\cite{acfkr} and that of the FSU5 models in ref. \cite{frs, hasan}. 
The standard--like models incorporate both the PS and FSU5 
breaking patterns and therefore include the basis vector
that breaks the $SO(10)$ symmetry to the PS subgroup, 
as well as a basis vector that breaks it to the FSU5
subgroup. The inclusion of two $SO(10)$ breaking vectors
is a unique characteristic of the SLMs, and impacts 
the space of vacua, as compared to the previous two cases. 
One reason is that each one of the $SO(10)$ breaking 
projections truncates the number of $SO(10)$ component states 
by two. Hence to produce three complete PS and FSU5 generations
requires that we start with an $SO(10)$ vacuum with a net number
of 6 generations, whereas the SLM models require an $SO(10)$ 
vacuum with 12 generations, and will severely restrict the 
number of SLM models with three complete generations.

\subsection{\emph{The Standard-like Construction}}

To construct the standard--like heterotic--string models in the free fermionic
formulation we therefore need to specify two additional basis vectors. 
The basis vector that breaks the $SO(10)$ symmetry to the PS subgroup 
can generically take the form
\begin{equation}
v_{13}=\alpha = \{ \overline{\psi}^{4,5}, \overline{\phi}^{1,2} \},
\label{alphav}
\end{equation} 
All other 
possible assignments that break the $SO(10)$ symmetry to the 
$SO(6)\times SO(4)$ are equivalent \cite{acfkr}.
Similarly to other free fermionic FSU5 and SLMs
constructed to date \cite{fsu5, slm, frs, hasan},
we restrict the assignment of 
rational phases of complex fermions to positive
1/2 boundary conditions.  
The choice of the FSU5 breaking vector $\beta$ is, however, not unique. 
The different choices were discussed in ref. \cite{frs}.
The basis vector $\beta$ in our SLM classification is taken to be
\begin{equation}
v_{14}=\beta = 
\{
\overline{\psi}^{1,\dots,5}=\textstyle\frac{1}{2},
\overline{\eta}^{1,2,3}=\textstyle\frac{1}{2},
\overline{\phi}^{1,2}  = \textstyle\frac{1}{2}, 
\overline{\phi}^{3,4}  = 1,
\overline{\phi}^{5,6}  = \textstyle\frac{1}{2},
\overline{\phi}^{7}   =  1,
\overline{\phi}^{8}=0\,\},\label{betav}
\end{equation}
and an alternative choice is given by
\begin{equation}
v_{14}=\beta' = 
 \{ 
\overline{\psi}^{1,...,5}=\frac{1}{2},
\overline{\eta}^{1,2,3}=\textstyle\frac{1}{2},
\overline{\phi}^{1,...,8}=\frac{1}{2} \}.
\nonumber
\end{equation}
The first choice ensures that the basis set is linearly independent, 
whereas the second is not as in this case we obtain
${\bf 1} = S + \sum_{i = 1}^{6} e_i + 2 \beta$, 
which results in correlations among the GGSO phases.
Classification of the FSU5 models using the second choice was 
discussed in \cite{hasan}. Here we will focus 
on the first choice. Our basis therefore consists of a set of 
14 independent basis vectors, 
$\{ 1,S,e_1,e_2,e_3,e_4,e_5,e_6, b_1,b_2,z_1,z_2,\alpha, \beta \}$.



\subsection{\emph{GGSO Projections}}

The second ingredient needed to construct the string models are the GGSO 
projection coefficients that appear in the one--loop partition function,
$\cc{v_i}{v_j}$, spanning a $14\times 14$ matrix.
Only the terms with $i>j$ are
independent, and the remaining terms are fixed by modular invariance.
A priori there are therefore 92 independent coefficients corresponding
to $2^{92}$ string vacua.
We note that the use of rational boundary conditions in $\beta$
does not increase the number of possibilities because the 
product $\beta\cdot v_i$ fixes the phases mod $Z_2$, {\it i.e.}
to be either $\pm1$ or $\pm i$ but not both. 
Thirteen additional coefficients
are fixed by demanding that the models possess $N=1$ supersymmetry. 
Without loss of generality we impose the associated GGSO projection
coefficients
\ba
\cc{1}{1} = \cc{S}{1} = \cc{S}{e_{i}} = \cc{S}{b_{m}} =
\cc{S}{z_n} = \cc{S}{\alpha}  = \cc{S}{\beta} = -1 , &&\\
i=1,...,6, \, m = 1,2 , \, n = 1,2. &&\nn
\ea
leaving 79 independent coefficients, which we choose to be
\begin{eqnarray}
\cc{1}{1}, ~~~~~~~~~~
&&\cc{e_i}{b_A}, \ \ 
\cc{e_i}{z_n}, \ \ 
\cc{e_i}{\alpha},  \ \ 
\cc{e_i}{\beta}, \nn\\
\cc{e_i}{e_j}, ~i\ge j~,& & 
\cc{1}{b_A}, \ \ 
\cc{1}{z_A}, \ \ 
\cc{1}{\alpha}, \ \ 
\cc{\beta}{1},
\label{igso}\\
& &
\cc{b_1}{b_2}, \ \ 
\cc{b_A}{z_n}, \ \
 \cc{b_A}{\alpha}, \ \ 
 \cc{\beta}{b_A}, \nn \\ 
\cc{\alpha}{\beta},~~~~~~~~~& & 
\cc{z_1}{z_2}, \ \
\cc{z_A}{\alpha}, \ \ 
\cc{\beta}{z_1}, \ \ \cc{z_2}{\beta}, \ \ 
\nn\\
i,j=1,\dots6\,\ ,\  A,n=1,2\nn.
\end{eqnarray}
All $\cc{i}{j}$ above are  real and take values $\pm1$.




\section{\emph{The String Spectrum}}\label{analysisspec}

As in previous cases we derive algebraic conditions for the
Generalised GSO (GGSO) projections on all the sectors that can produce 
massless states in the string standard--like models (SLMs). 
We remark here that the nomenclature ``standard--like models'' 
refers in this paper, and in conformity with earlier literature \cite{slm}, 
to the case in which the non--Abelian $SO(10)$ symmetry is reduced at 
the string level
to the non--Abelian subgroup $SU(3)\times SU(2)$ times the Abelian
subgroup $U(1)^2$. As the Standard Model contains a single Abelian 
group, this entails that the SLM models contain an additional 
Abelian group, beyond the Standard Model, that has to be broken
in the effective field theory limit. This point is particularly 
relevant to the exemplary model that we will present in section
\ref{example} and the Higgs states in the string SLM models that are 
available for breaking the additional Abelian symmetry. 
The algebraic constraints depend on the one loop 
GGSO phases and are coded in a computer program that scans the space
of vacua. However, due to the number of independent free phases 
we adopt a new strategy for extracting the phenomenologically interesting 
models. Whereas in the cases of the $SO(10)$ \cite{fknr}, \SO64 \cite{acfkr}
and \SU51 classifications \cite{frs, hasan}
the entire set of independent phases 
was spanned, in the case of the \S321 
models\footnote{We adopt $U(1)_C={3\over2}U(1)_{B-L};
U(1)_L=2U(1)_{T_{3_R}},$ in conformity with earlier literature.},
due to substantially larger number 
of choices, we adopt an alternative strategy. In the previous cases the entire
set of free phases for a string model was generated randomly and analysed
by imposing the GGSO projections in algebraic form. In the case
of the SLM vacua we generate a random choice of phases with unbroken 
$SO(10)$ symmetry and a net number of generations larger or equal to twelve, 
which is a minimal condition to generate three generation Standard--like 
Models. Additionally, we impose a set of conditions, to be discussed below,
on the randomly generated sets of GGSO projection phases that 
involves only the $SO(10)$ preserving basis vectors in eq. (\ref{basis}). 
We then perform a complete scan of the phases associated with the breaking 
of the $SO(10)$ symmetry down to the Standard Model subgroup. This method
generates a sizable space of three generation SLMs. We remark that the 
frequency 
of a three generation SLM is about one in $10^{12}$ and using the random 
generation 
of the entire set of free phases, in general, misses the phenomenologically 
viable cases. 

Similarly, to the previous cases the string states can be divided according 
to the 
sectors in which they arise, and algebraic conditions generated for the entire 
spectrum. Spacetime vector bosons arising in the untwisted sector generate 
the $SO(10)$ 
symmetry and its unbroken subgroups. 
The models contain additional sectors that may give rise to spacetime vector 
bosons and 
enhance the untwisted gauge symmetry. The twisted sectors in the models 
produce $N=1$ 
supersymmetric matter multiplets that may be classified according to the 
$SO(10)$ subgroup 
that they leave unbroken. Sectors that contain a linear combination of the 
basis vector 
$\alpha$ break the $SO(10)$ symmetry to the \SO64 subgroup, whereas sectors 
that contain a single combination with the basis vector $\pm\beta$ break the 
$SO(10)$ symmetry
to the FSU5 subgroup. Sectors that contain the combination $\alpha\pm\beta$ 
break the 
$SO(10)$ symmetry to the Standard Model subgroup. Sectors that contain the 
combination
$2\beta$ do not break the $SO(10)$ symmetry. All the remaining basis 
vectors do not break the
$SO(10)$ symmetry. Any sector that is obtained from combination of 
the $SO(10)$ preserving
vectors produces components of $SO(10)$ representations decomposed 
under the \S321 subgroup of
$SO(10)$, but that are not exotic with respect to the $U(1)$ 
Cartan generators of $SO(10)$, or that are $SO(10)$ singlets.
In contrast, the sectors that contain an $SO(10)$ breaking basis 
vector give rise to 
exotic states that carry exotic charges with respect to an 
unbroken $U(1)$ generator 
of the $SO(10)$ subgroup. The sectors that 
contain the \SO64 or \SU51 breaking vectors
produce states that carry fractional $U(1)_Y$ charge and hence 
fractional electric charge 
$\pm1/2$. Sectors that contain the combination $\alpha\pm\beta$ 
produce states that carry the 
standard charges under the Standard Model gauge group but carry 
fractional charges under the $U(1)_{Z^\prime}$ 
combination in eq. (\ref{u1zprime}). 

Additionally, the states producing sectors can be divided according to the 
left-- and right--moving vacuum. The physical states satisfy 
the Virasoro condition: 
\begin{equation}
M_L^2=-\frac{1}{2} + \frac{\xi_L\cdot\xi_L}{8}+N_L=
      -1           + \frac{\xi_R\cdot\xi_R}{8}+N_R=
M_R^2,
\label{virasorocon}
\end{equation}
where $N_L$ and $N_R$ are sums over the oscillators acting on the vacuum in the
left-- and right--moving sectors, respectively \cite{fff}. 
Sectors with $M_L=0$ and $M_R=0,4,6,8$ can produce spacetime vector bosons, 
which determine the gauge symmetry in a given vacuum configuration. 
Sectors with $M_L=4$ and $M_R=4,6,8$ produce matter states that will be 
enumerated below. All the models that we consider here preserve $N=1$
spacetime supersymmetry, which is generated by the single basis vector
$S$ with $(M_L;M_R)=(4,0)$.

\subsection{\emph{The gauge symmetry}}\label{gsymmetry}

The untwisted sector gives rise to spacetime vector bosons 
that correspond to the generators of the observable and hidden
sectors gauge symmetries
\begin{eqnarray}
{\rm Observable} &: &~~~~SU(3)_{C} \times U(1)_C\times SU(2)_{L}\times U(1)_L
                         \times U(1)_1 \times U(1)_2 \times U(1)_3 
\label{ogaugesym}\\
{\rm Hidden}     &: &~~~~SU(2)_{h1} \times U(1)_{h1} \times 
                         SU(2)_{h2} \times SU(2)_{h3}\times \nn\\
                 & & ~~~~ SU(2)_{h4} \times U(1)_{h4}\times 
                         U(1)_{h2}\times U(1))_{h3}             
\label{hgaugesym}
\end{eqnarray}

The $SO(10)$ symmetry breaking pattern is well known in 
Grand Unified Theories \cite{gutreviews, slm, asaka}.
The weak hypercharge is given by the combination 
\beq
U(1)_Y = \frac{1}{3} U(1)_C + \frac{1}{2} U(1)_L~,
\label{weakhyper}
\eeq
whereas the orthogonal $U(1)_{Z^\prime}$ combination is given by
\beq
U(1)_{Z^\prime} = U(1)_C  -  U(1)_L~.
\label{u1zprime}
\eeq

Depending on the choices of the GGSO projection coefficients,
additional massless spacetime vector bosons may be obtained from the 
following sectors,

\begin{equation}
\mathbf{G} =
\left\{ \begin{array}{cccc}
    x      &       z_1,& z_2,     & z_1 + z_2,  \\
x+2\beta, &   z_1+x+2\beta,   &  z_2+x+2\beta, &     z_1+z_2+x+2\beta,    \\
          &                  &                &                          \\
\alpha,   & \alpha+z_1,&\alpha+z_2, & \alpha+z_1+z_2, \\
\alpha+2\beta, & \alpha+x+z_1,& z_2+\alpha+2\beta,& z_1+z_2+x+\alpha+2\beta,\\
 \alpha+x, &  x+z_1+\alpha+2\beta &  x+\alpha+2\beta,   &  x+z_2+\alpha+2\beta    \\
          &  &     &     \\
\pm\beta, & z_1\pm\beta & z_2\pm\beta,& z_1+z_2\pm\beta,    \\
x\pm\beta &   x+z_1\pm\beta,& x+z_2\pm\beta,    & x+z_1+z_2\pm\beta,,    \\
         &  &     &    \\
\alpha\pm\beta &z_1+\alpha\pm\beta& z_2+\alpha\pm\beta& z_1+z_2+\alpha\pm\beta,  \\
x+\alpha\pm\beta,& x+z_1+\alpha\pm\beta& x+z_2+\alpha\pm\beta& x+z_1+z_2+\alpha\pm\beta \\
\end{array} \right\}. \label{ggsectors1}
\end{equation}
where 
\begin{equation}
x={\bf1}+S+\sum_{i=1}^6e_i+z_1+z_2=\{{\bar\psi}^{1,\cdots, 5},{\bar\eta}^{1,2,3}\}.
\label{xsector}
\end{equation}
There are in total 36 sectors that can produce massless spacetime vector bosons and hence 
enhance the gauge symmetry. The sectors in eq. (\ref{ggsectors1}) are divided 
according to the $SO(10)$ subgroup that they leave unbroken. The first two rows contain
sectors that do not break the $SO(10)$ symmetry, whereas rows 3--5, 6--7 and 8--9 break 
the $SO(10)$ symmetry to the $SO(6)\times SO(4)$, $SU(5)\times U(1)$ and 
$SU(3)\times SU(2)\times U(1)^2$, respectively. 

In the classification of the $SO(10)$ \cite{fknr}, \SO64 \cite{acfkr}
and the \SU51 \cite{frs} type of vacua the conditions for survival of 
vector bosons
from the enhancing sectors were derived. It was then imposed that all the 
spacetime vector bosons from these sectors are projected out. The gauge 
symmetry in these cases therefore only arose from the generators that are
obtained in the untwisted Neveu--Schwarz sector. In the case of the 
\S321 models, as seen from equation (\ref{ggsectors1}), the number of 
enhancing sectors proliferates, rendering the previous approach impractical. 
We therefore adopt an alternative strategy. The condition for projection 
of the enhanced symmetries are not derived. The space of scanned vacua
therefore contains models with enhanced symmetries, which amounts to 
about 20\% of the models. In extracting viable models we impose some
phenomenological constraints and restrict that these models only contain
enhancements of the hidden sector gauge group. 

\subsection{\emph{The Twisted Matter Sectors}}\label{twistedmatter}

\subsubsection{\emph{General Remarks}}

The proliferation of gauge symmetry enhancing sectors implies that there is 
a similar proliferation in the twisted sectors. The string models that we consider
correspond to $Z_2\times Z_2$ orbifolds, which contain three twisted sectors. 
The primary twisted sectors, or twisted planes, are generated by 
the vectors $b_1$, $b_2$ and $b_3=b_1+b_2+x$. Each twisted sector 
of the $Z_2\times Z_2$ orbifold contains sixteen fixed points, 
which we denote as $B_i^{pqrs}$ with $i=1,2,3$ denoting the twisted 
plane and $p,q,r,s=0,1$ denoting the fixed points. Since all
of the symmetry enhancing sectors in eq. (\ref{ggsectors1})
are blind to the internal twisted space, adding any of these
sectors to the primary sectors $b_{1,2,3}$ can produce physical 
massless states. If spacetime vector bosons from a given enhancing 
sector survive the GGSO projection, the states arising from its combination 
with the primary twisted sectors $b_{1,2,3}$ merely complements the
physical multiplets in that sector to representations of the 
enhanced symmetry. If the additional vector bosons are projected 
out, as is the case with respect to most of the symmetry enhancing sectors, 
then its combination with the primary twisted sectors will produce 
states that are singlets of the enhanced symmetry, but transform under 
other part of the four dimensional gauge group. For example, 
in models with unbroken $SO(10)$, the $x$--sector enhances the 
$SO(10)\times U(1)$ gauge symmetry to $E_6$. If the symmetry enhancing
states from the $x$--sector are projected out then the $10+1+1$ 
representations 
of $SO(10)$ that arise in the sectors $b_i+x$ are mapped to vectorial
representations of the hidden sector gauge group \cite{xmap}. 
We remark that the $10+1+1$ states correspond to the $10+1$ in the 
$27$ chiral representation of $E_6$, which decomposes under $SO(10$ 
as $16+10+1$, whereas the additional $E_6$ singlet correspond to a 
twisted moduli \cite{xmap, moduli}.
All the sectors arising in the twisted planes preserve the underlying
structure of a six dimensional $Z_2\times Z_2$ toroidal orbifold. 

The sectors in the string models can be further divided into those 
that do not break the $SO(10)$ symmetry and those that do.
Sectors that preserve the $SO(10)$ symmetry are divided into sectors 
that produce observable states, that transform under the Standard Model
gauge group, and sectors that produce hidden states that do not carry 
Standard Model charges. Sectors that break the $SO(10)$ symmetry 
are divided by the $SO(10)$ symmetry breaking pattern in each of the sectors. 
Additionally, the sectors are divided by the right--moving vacuum.
To produce massless states,
sectors with $\xi_R\cdot \xi_R=4$ require one right--moving 
NS oscillator acting on the vacuum, whereas 
sectors with $\xi_R\cdot \xi_R=6$ require one
oscillator of a worldsheet fermion with boundary condition $\pm1/2$. 
Sectors with $\xi_R\cdot \xi_R=8$ do not require any oscillators. 

\subsubsection{\emph{The Observable Matter Sectors}}\label{analysis2}

Similarly to the cases of the PS and FSU5 models, the observable matter
spectrum 
arises from sectors that leave the underlying $SO(10)$ symmetry unbroken. The
observable matter states therefore arise from $SO(10)$ representations, 
decomposed under the \S321 subgroup. The Standard Model states may arise from 
spinorial or vectorial $SO(10)$ representations. Additionally, these sectors 
may give rise to states that are $SO(10)$, and consequently, SLM singlets. 
The chiral spinorial representations of the observable \S321 
arise from the sectors:

\begin{eqnarray} \label{obspin}
B_{pqrs}^{(1)}&=& S + {b_1 + p e_3+ q e_4 + r e_5 + s e_6} \nonumber\\
&=&\{\psi^\mu,\chi^{12},(1-p)y^{3}\overline{y}^3,p\omega^{3}\overline{\omega}^3,
(1-q)y^{4}\overline{y}^4,q\omega^{4}\overline{\omega}^4, \nonumber\\
& & ~~~(1-r)y^{5}\overline{y}^5,r\omega^{5}\overline{\omega}^5,
(1-s)y^{6}\overline{y}^6,s\omega^{6}\overline{\omega}^6,
\overline{\eta}^1,\overline{\psi}^{1,...,5}\},
\\
B_{pqrs}^{(2)}&=& S + {b_2 + p e_1 + q e_2 + r e_5 + s e_6},
\label{twochiralspinorials}
\nonumber\\
B_{pqrs}^{(3)}&=& S + {b_3 + p e_1+ q e_2 + r e_3 + s e_4}, \nonumber
\end{eqnarray}
where $p,q,r,s=0,1$ and $b_3=b_1+b_2+x$.
These 48 sectors give rise to $\textbf{16}$ and $\overline{\textbf{16}}$ 
multiplets of $SO(10)$ decomposed under
$SU(3) \times U(1)_C \times SU(2)\times U(1)_L$, which are given by
\begin{eqnarray}
\textbf{16} &=&  ~~
\left(\textbf{3},+\textstyle\frac{1}{2};\textbf{2},0\right)\oplus
\left(\overline{\textbf{3}},-\textstyle\frac{1}{2};\textbf{1},+1\right)\oplus
\left(\overline{\textbf{3}},-\textstyle\frac{1}{2};\textbf{1},-1\right)
\label{sms}\\
&&\oplus
\left(\textbf{1}, -\textstyle\frac{3}{2}; \textbf{2}, 0\right)\oplus
\left(\textbf{1}, +\textstyle\frac{3}{2}; \textbf{1}, +1 \right)\oplus
\left(\textbf{1}, +\textstyle\frac{3}{2}; \textbf{1}, -1\right)
\nonumber\\
& & \nn\\
\overline{\textbf{16}} &=& ~~
\left(\overline{\textbf{3}},-\textstyle\frac{1}{2};\textbf{2},0\right)\oplus
\left(\textbf{3},+\textstyle\frac{1}{2};\textbf{1},-1\right)\oplus
\left(\textbf{3},+\textstyle\frac{1}{2};\textbf{1},+1\right)
\label{smsb}\\
&&\oplus\left(\textbf{1}, +\textstyle\frac{3}{2}; \textbf{2}, 0\right)\oplus
\left(\textbf{1}, -\textstyle\frac{3}{2}; \textbf{1}, -1 \right)\oplus
\left(\textbf{1}, -\textstyle\frac{3}{2}; \textbf{1}, +1\right).\nonumber
\end{eqnarray}
Additionally, vector--like representations of the observable
$SU(3) \times U(1)_C \times SU(2)\times U(1)_L$ 
gauge group arise from the sectors
\begin{eqnarray}
B_{pqrs}^{(4)}&=&B_{pqrs}^{(1)} + x \nonumber\\
&=& S + b_1 + p e_3+ q e_4 + r e_5 + s e_6 + x\nonumber\\
&=&\{\psi^\mu,\chi^{12},(1-p)y^{3}\overline{y}^3,p\omega^{3}\overline{\omega}^3,
(1-q)y^{4}\overline{y}^4,q\omega^{4}\overline{\omega}^4, \nonumber\\
& & ~~~(1-r)y^{5}\overline{y}^5,r\omega^{5}\overline{\omega}^5,
(1-s)y^{6}\overline{y}^6,s\omega^{6}\overline{\omega}^6,\overline{\eta}^{2,3} \}
\label{nonchiralvectorials}
\\
B_{pqrs}^{(5,6)}&=& B_{pqrs}^{(2,3)} +x.\nonumber
\end{eqnarray}
Massless states in these sectors are obtained by acting on the vacuum with a NS right--moving 
oscillator. They produce vectorial 10 representations of $SO(10)$ decomposed as
\begin{eqnarray}
\textbf{10} &= &
\left({\textbf{1}}, 0; {\textbf{2}}, +1\right)\oplus 
\left({\textbf{1}}, 0; {\textbf{2}}, -1\right) \oplus 
\left({\textbf{3}}, -1; {\textbf{1}}, 0\right) \oplus 
\left({\overline{\textbf{3}}}, +1; {\textbf{1}}, 0\right)  , 
\nonumber
\end{eqnarray}
where the electroweak doublet representations may be identified as light Higgs states. 
The sectors $B^{(4,5,6)}_{pqrs}$ may additionally produce the $SO(10)$ singlet states 

\begin{itemize}
\item 
$\{\overline\eta^{i}\}|R \rangle_{pqrs}^{(4,5,6)}$ or
$\{\overline\eta^{*i}\}|R\rangle_{pqrs}^{(4,5,6)}$, $i = 1,2,3$,
where $|R\rangle_{pqrs}^{(4,5,6)}$ is the degenerate Ramond vacuum of the
$B_{pqrs}^{(4,5,6)}$ sector.
These states transform as a vector--like representations under the $U(1)_i$'s.

\item 
$\{\overline\phi^{j}\}|R\rangle_{pqrs}^{(4,5,6)}$ or
$\{\overline\phi^{*1,2}\}|R\rangle_{pqrs}^{(4,5,6)}$, $j=1, \dots, 8$.
These states transform as a vector--like representations of the hidden sector gauge group. 

\item 
$\{\overline{y}^{1,...,6}\}|R\rangle_{pqrs}^{(4,5,6)}$ or
$\{\overline\omega^{*1,...,6}\}|R\rangle_{pqrs}^{(4,5,6)}$.
These states transform as a vector--like representations under the $U(1)_i$'s.
\end{itemize}

We note that the states arising from the sectors in eqs.
(\ref{obspin}) and (\ref{nonchiralvectorials})
transform as standard states under the 
Standard Model gauge group. The term ``exotic 
states'' is reserved to states that carry 
non--standard charges with respect to the 
$SO(10)$ group factors. This distinction is particularly important in the
case of the SLMs. Exotic states in the SLMs are obtained from sectors that 
break the $SO(10)$ symmetry, {\it i.e.} sectors that contain the vectors 
$\alpha$, $\beta$ or their combination $\alpha\pm\beta$. However, while the 
first two cases carry fractional electric charges, the last category carry
non--standard $SO(10)$ charges with respect to $U(1)_{Z^\prime}$ rather than
with respect to the Standard Model subgroup.


The number of $SO(10)$ spinorials/anti-spinorials, $N_{16}/N_{\overline{16}}$, arising from the sectors $B_{pqrs}^{(A)}, A=1,2,3\,,$ $p,q, r,s=0,1\,,$
is determined by the projectors
\begin{eqnarray}
P^1_{pqrs}=
\frac{1}{2^4}\prod_{i=1,2}
\left(1-{\cc{B_{pqrs}^{(1)}}{e_1}}^\ast\right)
\prod_{a=1,2}
\left(1-{\cc{B_{pqrs}^{(1)}}{z_a}}^\ast\right)
\\
P^2_{pqrs}=
\frac{1}{2^4}\prod_{i=3,4}
\left(1-{\cc{B_{pqrs}^{(2)}}{e_i}}^\ast\right)
\prod_{a=1,2}
\left(1-{\cc{B_{pqrs}^{(2)}}{z_a}}^\ast\right)
\\
P^3_{pqrs}=
\frac{1}{2^4}\prod_{i=5,6}
\left(1-{\cc{B_{pqrs}^{(3)}}{e_i}}^\ast\right)
\prod_{a=1,2}
\left(1-{\cc{B_{pqrs}^{(3)}}{z_a}}^\ast\right)
\end{eqnarray}
and the phases
\begin{eqnarray}
X^1_{pqrs} & = &
-{\cc{B_{pqrs}^{(1)}}{S+b_2+(1-r)e_5+(1-s)e_6}}^\ast
\\
X^2_{pqrs} & = &
-{\cc{B_{pqrs}^{(2)}}{S+b_1+(1-r)e_5+(1-s)e_6}}^\ast
\\
X^3_{pqrs} & = &
-{\cc{B_{pqrs}^{(3)}}{S+b_1+(1-r)e_3+(1-s)e_4}}^\ast,
\end{eqnarray}
as follows
\begin{eqnarray}
N_{16} & = & \frac{1}{2}\sum_{\substack{A=1,2,3\\ p,q,r,s=0,1}} P^A_{pqrs}\left(1+X^A_{pqrs}\right) \label{spinor}\\
N_{\overline{16}}& = & \frac{1}{2}\sum_{\substack{A=1,2,3\\ p,q,r,s=0,1}} P^A_{pqrs}\left(1-X^A_{pqrs}\right)\ . \label{aspinor}
\end{eqnarray}
Here we have assumed the chirality of the spacetime fermions to be $c\left(\psi^\mu\right)=+1$.

Similarly, the number of $SO(10)$ vectorials, $N_{10}$, is determined by the projectors
\begin{eqnarray}
R^1_{pqrs}=
\frac{1}{2^4}\prod_{i=1,2}
\left(1-{\cc{B_{pqrs}^{(1)}+x}{e_i}}^\ast\right)
\prod_{a=1,2}
\left(1-{\cc{B_{pqrs}^{(1)}+x}{z_a}}^\ast\right)
\\
R^2_{pqrs}=
\frac{1}{2^4}\prod_{i=3,4}
\left(1-{\cc{B_{pqrs}^{(2)}+x}{e_i}}^\ast\right)
\prod_{a=1,2}
\left(1-{\cc{B_{pqrs}^{(2)}+x}{z_a}}^\ast\right)
\\
R^3_{pqrs}=
\frac{1}{2^4}\prod_{i=5,6}
\left(1-{\cc{B_{pqrs}^{(3)}+x}{e_i}}^\ast\right)
\prod_{a=1,2}
\left(1-{\cc{B_{pqrs}^{(3)}+x}{z_a}}^\ast\right)
\end{eqnarray}
as follows
\begin{eqnarray}
N_{10}=\sum_{\substack{A=1,2,3\\ p,q,r,s=0,1}} R^A_{pqrs} .
\label{vector}
\end{eqnarray}

Furthermore, after applying the $\alpha, \beta$ projections onto the
remaining spinorials/vectorials in order to obtain the final SM states, 
we observe that some of these $SO(10)$ spinorials/vectorials are
entirely  projected out. 
A detailed analysis shows that the surviving SM states originate  from specific $SO(10)$ spinorials/vectorials that satisfy certain criteria that can be
expressed in terms of GGSO phases involving only the basis vectors \eqref{basis}.
Utilising the following projectors
\begin{eqnarray}
S^I_{pqrs}&=&
-{\cc{B_{pqrs}^{(I)}}{x}}^\ast\ ,\ I=1,2,3\\
T^I_{pqrs}&=&
-{\cc{B_{pqrs}^{(I+3)}}{x}}^\ast\ \ ,
\end{eqnarray}
we can demonstrate
\footnote{The easiest way to verify this is to consider the projection of the
vector $x+2\beta$ onto $B^I_{pqrs}$ taking into account that
$x+2\beta=\left\{\bar{\phi}^{1,2,5,6}\right\}\cap B^I_{pqrs}=\varnothing$ and 
$\cc{B^I_{pqrs}}{x+2\beta}=-\cc{B^I_{pqrs}}{x}$.}
 that the surviving SM states in $B^I_{pqrs}$ and 
$B^{I+3}_{pqrs},\,I=1,2,3$ arise solely from spinorials/vectorials with
\begin{eqnarray}
S^I_{pqrs}=-1\ ,\ T^I_{pqrs}=-1\ ,\ I=1,2,3
\label{fcon}
\end{eqnarray}
respectively.
The number of these fertile  $\textbf{16}/\overline{\textbf{16}}$ and $\textbf{10}$ can be also expressed exclusively in terms of $SO(10)$ level projectors, that is GGSO coefficients involving only the first 12 basis vectors, as follows
\begin{eqnarray}
N^f_{16}&=&\frac{1}{2}\sum_{\substack{A=1,2,3\\ p,q,r,s=0,1}} P^A_{pqrs}S^A_{pqrs}\left(1+X^A_{pqrs}\right)\label{nss}\\
N^f_{\overline{16}}&=&\frac{1}{2}\sum_{\substack{A=1,2,3\\ p,q,r,s=0,1}} P^A_{pqrs}S^A_{pqrs}\left(1-X^A_{pqrs}\right)\label{nsbs}\\
N^f_{10}&=&\sum_{\substack{A=1,2,3\\ p,q,r,s=0,1}} R^A_{pqrs} T^A_{pqrs} .
\label{nvs}
\end{eqnarray}
These expressions can be further analysed and written in terms of the GGSO coefficients of \eqref{igso}. 
After some algebra we arrive to the conclusion that the number of independent $\cc{i}{j}$ involved is 44. These are
\begin{eqnarray}
\cc{1}{1},\ \cc{1}{b_A},\ \cc{e_i}{e_j}_{i<j},\ \cc{e_i}{z_n},\ 
 \cc{b_A}{z_n},\ \cc{b_1}{b_2},\ \cc{z_1}{z_2}\nonumber\\
 \cc{e_1}{b_1},\ \cc{e_2}{b_1},\ \cc{e_5}{b_1},\ \cc{e_6}{b_1},\ 
 \cc{e_3}{b_2},\ \cc{e_4}{b_2},\ \cc{e_5}{b_2},\ \cc{e_6}{b_2},\ 
 \label{parama}
\end{eqnarray}
where without loss of generality, as far as the
spinorial/vectorial and descendant states are concerned,
we have assumed
\begin{eqnarray}
\cc{1}{e_i}=\cc{1}{z_n}=\cc{e_3}{b_1}=\cc{e_4}{b_1}=\cc{e_1}{b_2}=
\cc{e_2}{b_2}=+1\ .
\label{ggsoca}
\end{eqnarray}

After having identified the fertile $SO(10)$ spinorials/vectorials we 
turn to the explicit application of the remaining projections related 
to $\alpha$ and $\beta$ vectors.
As explained earlier, for generic points of the parameter space, 
these projections break $SO(10)$ to \S321 and truncate the fertile 
$SO(10)$ spinorials/vectorials. 
The surviving states and their SM content for the various choices 
of the projectors are shown in table \ref{ptable} for the case of 
$SO(10)$ spinorials/antispinorials and in 
table \ref{ptablev} for the case of $SO(10)$ vectorials, where
\begin{eqnarray}
U^{\alpha,I}_{pqrs}& = &
-{\cc{B^I_{pqrs}}{\alpha}}^\ast\ ~~~~~~~~,\ 
~~U^{\beta,I}_{pqrs}=
-{\cc{B^I_{pqrs}}{\beta}}^\ast\ , \\
V^{\alpha,I}_{pqrs}& = &
-{\cc{B^I_{pqrs}+x}{\alpha}}^\ast\ ~~ ,\ ~~
V^{\beta,I}_{pqrs}=
-{\cc{B^I_{pqrs}+x}{\beta}}^\ast\ ,
\end{eqnarray}
and
\begin{eqnarray}
Y^1_{pqrs}& = &
-{\cc{B_{pqrs}^{(1)}+x}{S+b_2+(1-r)e_5+(1-s)e_6}}^\ast
\\
Y^2_{pqrs}& = &
-{\cc{B_{pqrs}^{(2)}+x}{S+b_1+(1-r)e_5+(1-s)e_6}}^\ast
\\
Y^3_{pqrs}& = &
-{\cc{B_{pqrs}^{(3)}+x}{S+b_1+(1-r)e_3+(1-s)e_4}}^\ast\ .
\end{eqnarray}
Here ${d^c}', \overline{{d^c}'}$ are additional SM triplet pairs and $\overline{Q}, \overline{u^c},\overline{e^c}, \overline{d^c},\overline{\nu^c}$ are additional states
with quantum numbers conjugate to those of corresponding SM states.

\begin{table}[h]
\begin{center}
\begin{tabular}{ccccc}
\hline
$X^I_{pqrs}$ & $U^{\alpha,I}_{pqrs}$ & $U^{\beta,I}_{pqrs}$ & \S321 rep(s) & SM state(s)\\
\hline
$+1$&$+1$&$+i$&$\left(\overline{\textbf{3}},-\textstyle\frac{1}{2};\textbf{1},-1\right)\oplus \left(\textbf{1}, +\textstyle\frac{3}{2}; \textbf{1}, +1 \right)$&$u^c,e^c$\\
\hline
$+1$&$+1$&$-i$&$\left(\overline{\textbf{3}},-\textstyle\frac{1}{2};\textbf{1},+1\right)\oplus \left(\textbf{1}, +\textstyle\frac{3}{2}; \textbf{1}, -1 \right)$&$d^c,\nu^c$\\
\hline
$+1$&$-1$&$+i$&$\left(\textbf{1}, -\textstyle\frac{3}{2}; \textbf{2}, 0\right)$&$L$\\
\hline
$+1$&$-1$&$-i$&$\left(\textbf{3},+\textstyle\frac{1}{2};\textbf{2},0\right)$&$Q$\\
\hline
$-1$&$+1$&$+i$&$\left({\textbf{3}},+\textstyle\frac{1}{2};\textbf{1},+1\right)\oplus \left(\textbf{1}, -\textstyle\frac{3}{2}; \textbf{1}, -1 \right)$&$\overline{u^c},\overline{e^c}$\\
\hline
$-1$&$+1$&$-i$&$\left({\textbf{3}},+\textstyle\frac{1}{2};\textbf{1},-1\right)\oplus \left(\textbf{1}, -\textstyle\frac{3}{2}; \textbf{1}, +1 \right)$&$\overline{d^c},\overline{\nu^c}$\\
\hline
$-1$&$-1$&$+i$&$\left(\textbf{1}, +\textstyle\frac{3}{2}; \textbf{2}, 0\right)$&$\overline{L}$\\
\hline
$-1$&$-1$&$-i$&$\left(\overline{\textbf{3}},-\textstyle\frac{1}{2};\textbf{2},0\right)$&$\overline{Q}$\\
\hline
\end{tabular}
\end{center}
\caption{\label{ptable}
Offspring states arising from fertile $SO(10)$ spinorials/antispinorials 
($X^I_{pqrs}=+1/-1)$, for the allowed values of the projectors 
$U^{\alpha,I}_{pqrs}$,  $U^{\beta,I}_{pqrs}$ related to the $SO(10)$
 breaking vectors  $\alpha, \beta$,  and their transformation 
under \S321. The related SM states, in the case of standard 
Hypercharge embedding, are shown in the last column.}
\end{table}

A detailed analysis of the additional GGSO projectors presented above 
shows that as far as offspring spinorial and vectorial $SO(10)$ states 
are concerned we have 18 additional independent phases
involved. These are  
\begin{eqnarray}
\cc{1}{\alpha},\  \cc{\beta}{1},\ \cc{e_i}{\beta}\ ,\cc{e_i}{\alpha} ,\ 
\cc{b_a}{\alpha},\ \cc{\beta}{b_a}\ ,\ {i=1\dots6}, {a=1,2}
\label{parame}
\end{eqnarray}
where we have appropriately chosen as parameters the phases
with allowed values $\pm1$. Moreover, without loss of generality 
for the states under consideration
we have set 
\begin{eqnarray}
\cc{z_1}{\alpha}=\cc{z_2}{\alpha}=\cc{\beta}{z_1}=\cc{\beta}{z_2}=+1\ .
\end{eqnarray}
Using the information presented above we can calculate for each model 
in this class the following numbers $n(Q), n(\overline{Q}), n(L), 
n(\overline{L}), n(d^c), n(\overline{d^c}), 
n(u^c), n(\overline{u^c}), n({d^c}'), n(\overline{{d^c}'})$, $n(H_u)$, 
$n(H_d)$, corresponding to the multiplicities of the associated SM fields, 
in terms of the 44+18=62 independent GGSO phases. 
Of course, in realistic cases these numbers are not 
independent.  For example, a minimal set of phenomenological requirements 
includes: 
\begin{itemize}
\item{(i)}
Complete fermion generations, that is 
\begin{eqnarray}
n(Q)-n(\overline{Q})= 
n(L)-n(\overline{L})=
n(d^c)-n(\overline{d^c})=
n(u^c)-n(\overline{u^c})=n_g\ ,
\label{famco}
\end{eqnarray}
where $n_g$ the generation number.
\item
(ii)Absence of mixed states transforming both under the SM and some 
hidden sector non Abelian gauge group factors. This requires additional 
states to appear in vector-like
pairs, otherwise cancellation of mixed anomalies infers the presence of 
states in mixed representations. So in addition to (i) we have to impose 
\begin{eqnarray}
n(H_u)=n(H_d)=n_H\ ,\ n({d^c}')=n(\overline{{d^c}'})= n_d
\label{damco}
\end{eqnarray}
\item
(iii) Existence of SM breaking Higgs doublets, that is
\begin{eqnarray}
n_H\ge1\ .
\label{hamco}
\end{eqnarray}
\end{itemize}

\begin{table}[h]
\begin{center}
\begin{tabular}{ccccc}
\hline
$Y^I_{pqrs}$ & $V^{\alpha,I}_{pqrs}$ & $V^{\beta,I}_{pqrs}$ & \S321 rep(s) 
                      & SM state(s)\\
\hline
$\pm1$&$-1$&$\pm i$&$\left({\textbf{1}}, 0; {\textbf{2}}, +1\right)$&$H_u$\\
\hline
$\pm1$&$-1$&$\mp i$&$\left({\textbf{1}}, 0; {\textbf{2}}, -1\right)$&$H_d$\\
\hline
$\pm1$&$+1$&$\pm i$&$\left({\textbf{3}}, -1; {\textbf{1}}, 0\right)$&$\overline{{d^c}'}$\\
\hline
$\pm1$&$+1$&$\pm i$&$\left({\overline{\textbf{3}}}, +1; {\textbf{1}}, 0\right)$&${d^c}'$\\
\hline
\end{tabular}
\end{center}
\caption{\label{ptablev}
Offspring  states arising from fertile $SO(10)$ vectorials, for the allowed values 
of the projectors $Y^I_{pqrs}$, $V^{\alpha,I}_{pqrs}$, $V^{\beta,I}_{pqrs}$,
and their transformation under \S321.  In the last column we present the related 
SM states in the case of standard Hypercharge embedding.
}
\end{table}
 
\subsubsection{\emph{The Hidden Matter Sectors}}\label{analysis3}

The hidden matter spectrum arises in sectors that do not break the $SO(10)$ symmetry 
but that do not transform under the \S321 subgroup of $SO(10)$. All the 
sectors in this category have $\xi_R\cdot \xi_R= 8$. They may contain the combination
of the $SO(10)$ preserving vectors $x$, $z_{1,2}$ and $2\beta$, but not of the 
$SO(10)$ breaking vectors $\alpha$ and $\beta$. The $SO(10)$ preserving hidden 
sectors are: 
\begin{eqnarray}\label{hidSO10}
B_{pqrs}^{(7,8,9)} ~~~&=& B_{pqrs}^{(1,2,3)} + 2\beta \nonumber\\
B_{pqrs}^{(10, 11,12)} &=& B_{pqrs}^{(1,2, 3)} + 2\beta+ z_1 \nonumber\\
B_{pqrs}^{(13, 14,15)} &=& B_{pqrs}^{(1,2, 3)} + 2\beta+ z_2 \nonumber\\
B_{pqrs}^{(16, 17,18)} &=& B_{pqrs}^{(1,2, 3)} +2\beta+ z_1+ z_2 \\
B_{pqrs}^{(16, 17,18)} &=& B_{pqrs}^{(1,2, 3)} +x+ z_1 \nonumber\\
B_{pqrs}^{(16, 17,18)} &=& B_{pqrs}^{(1,2, 3)} +x+ z_2 \nonumber
\end{eqnarray}

%

\subsubsection{\emph{The Exotic Matter Sectors}}\label{analysis4}

Exotic matter sectors arise from combinations of the $SO(10)$ basis vectors
$\alpha$ and $\pm\beta$ with the other basis vectors. These sectors 
can be divided according to the $SO(10)$ subgroup that they leave unbroken. 
There are three possibilities:
\begin{itemize} 
\item 1. the vector $\alpha$ leaves the \SO64 subgroup unbroken; 
\item 2. the vector $\pm\beta$ leaves the \SU51 subgroup unbroken;
\item 3. the combination $\alpha\pm\beta$ leaves \S321 unbroken. 
\end{itemize} 
Additionally, the sectors can be divided by the right--moving product
$\xi_R\cdot\xi_R = 6, 8$, where the first case can only leave 
unbroken the \SU51 or \S321 subgroups but not the \SO64 one. That
is, this case must include the basis vector $\pm\beta$. 
The \SU51 preserving sectors are: 

\begin{itemize}
\item $\xi_R\cdot \xi_R=6$ 

\begin{eqnarray}\label{exo6su51}
B_{pqrs}^{(19, 20,21)} &=& B_{pqrs}^{(1,2, 3)} \pm \beta + z_1 \nonumber\\
B_{pqrs}^{(22, 23,24)} &=& B_{pqrs}^{(1,2, 3)} \pm \beta + z_1 +z_2\nonumber\\
B_{pqrs}^{(25, 26,27)} &=& B_{pqrs}^{(1,2, 3)} \pm \beta + x + z_1 \\
B_{pqrs}^{(28, 29,30)} &=& B_{pqrs}^{(1,2, 3)} \pm \beta + x + z_1 + z_2 \nonumber
\end{eqnarray}
\item $\xi_R\cdot \xi_R=8$ 
\begin{eqnarray}\label{exo8su51}
B_{pqrs}^{(31, 32,33)} &=& B_{pqrs}^{(1,2, 3)} \pm \beta       \nonumber\\
B_{pqrs}^{(34, 35,36)} &=& B_{pqrs}^{(1,2, 3)} \pm \beta + z_2\nonumber\\
B_{pqrs}^{(37, 38,39)} &=& B_{pqrs}^{(1,2, 3)} \pm \beta + x  \\
B_{pqrs}^{(40, 41,42)} &=& B_{pqrs}^{(1,2, 3)} \pm \beta + x + z_2 \nonumber
\end{eqnarray}
\end{itemize}

The \SO64 preserving sectors can only have $\xi_R\cdot \xi_R=8$. These are:
\begin{eqnarray}\label{exo8so64}
B_{pqrs}^{(43, 44,45)} &=& B_{pqrs}^{(1,2, 3)} + \alpha       \nonumber\\
B_{pqrs}^{(46, 47,48)} &=& B_{pqrs}^{(1,2, 3)} + \alpha + 2 \beta\nonumber\\
B_{pqrs}^{(49, 50,51)} &=& B_{pqrs}^{(1,2, 3)} + \alpha + z_1  \nonumber\\
B_{pqrs}^{(52, 53,54)} &=& B_{pqrs}^{(1,2, 3)} + \alpha + z_2 +2 \beta \\
B_{pqrs}^{(55, 56,57)} &=& B_{pqrs}^{(1,2, 3)} + \alpha + x \nonumber\\
B_{pqrs}^{(58, 59,60)} &=& B_{pqrs}^{(1,2, 3)} + \alpha + x + z_1 \nonumber\\
B_{pqrs}^{(61, 62,63)} &=& B_{pqrs}^{(1,2, 3)} + \alpha + x + 2 \beta\nonumber\\
B_{pqrs}^{(64, 65,66)} &=& B_{pqrs}^{(1,2, 3)} + \alpha + x + z_2 +2 \beta\nonumber
\end{eqnarray}

Finally, the \S321 preserving sectors are: 
\begin{itemize}
\item $\xi_R\cdot \xi_R=6$ 

\begin{eqnarray}\label{exo6su3211}
B_{pqrs}^{(67, 68,69)} &=& B_{pqrs}^{(1,2, 3)} \alpha\pm \beta + z_1 \nonumber\\
B_{pqrs}^{(70, 71,72)} &=& B_{pqrs}^{(1,2, 3)} \alpha\pm \beta + z_1 +z_2\nonumber\\
B_{pqrs}^{(73, 74,75)} &=& B_{pqrs}^{(1,2, 3)} \alpha\pm \beta + x + z_1 \\
B_{pqrs}^{(76, 77,78)} &=& B_{pqrs}^{(1,2, 3)} \alpha\pm \beta + x + z_1 + z_2 \nonumber
\end{eqnarray}
\item $\xi_R\cdot \xi_R=8$ 
\begin{eqnarray}\label{exo8su3211}
B_{pqrs}^{(79, 80,81)} &=& B_{pqrs}^{(1,2, 3)} \alpha\pm \beta       \nonumber\\
B_{pqrs}^{(82, 83,84)} &=& B_{pqrs}^{(1,2, 3)} \alpha\pm \beta + z_2\nonumber\\
B_{pqrs}^{(85, 86,87)} &=& B_{pqrs}^{(1,2, 3)} \alpha\pm \beta + x  \\
B_{pqrs}^{(88, 89,90)} &=& B_{pqrs}^{(1,2, 3)} \alpha\pm \beta + x + z_2 \nonumber
\end{eqnarray}
\end{itemize}



%
\section{\emph{Top quark Yukawa coupling}}\label{tmy}
Apart from the SM spectrum, 
string theory is expected to reproduce also the SM 
interactions at the low energy limit. To verify this we need information
regarding the effective superpotential that  usually infers lengthy 
calculations of model 
dependent string amplitudes. 
However, it has been shown that for
the calculation of fermion mass terms, the related 
superpotential couplings can be 
implemented using a 
straightforward general analytical method \cite{topyukawa}.
Especially when applied to the top quark Yukawa coupling, which is 
in general expected to be present at the tree--level superpotential, 
the necessary conditions can be expressed in terms of GGSO phases.
We note here several 
distinctions between the early SLM constructions \cite{slm}
and the type of models that we analyse herein, which 
are particularly relevant for the top quark Yukawa coupling.
The key difference is that the early SLM constructions utilised asymmetric 
boundary conditions with respect to the set of worldsheet fermions 
$\{y,\omega\vert{\bar y}, {\bar\omega}\}^{1,\cdots,6}$, whereas 
the class of models that we consider here utilise symmetric boundary 
conditions. This results in the retention of untwisted electroweak 
doublets in the asymmetric SLM models \cite{slm, dts}, and their
projection in the symmetric SLM models considered here. The top quark Yukawa
coupling in the asymmetric SLM models arises therefore from a cubic
level coupling of twisted--twisted-untwisted string states \cite{slm, top}, 
whereas in the symmetric models it is obtained from a
twisted--twisted--twisted coupling of string states. 
In the asymmetric models the coupling is determined in terms of 
the boundary condition assignments \cite{top}, whereas in the 
symmetric models it constrains the GGSO phase assignments \cite{topyukawa}.
In the class of models under consideration the top 
quark mass coupling reads
$$
\lambda_t Q^F {u^{c}}^F H_u^B
$$
where the superscripts $F,B$ refer to the fermionic/bosonic
component of the associated superfield. 
As was shown in  \cite{topyukawa} the necessary conditions
for the presence of this coupling, before the breaking of the  
$SO(10)$ gauge symmetry, are
\begin{eqnarray}
\cc{b_1}{e_1}=\cc{b_1}{e_2}=\cc{b_2}{e_3}=\cc{b_2}{e_4}=
\cc{b_1}{z_1}=\cc{b_1}{z_2}=
\cc{b_2}{z_1}=\cc{b_2}{z_2}=+1 \ ,\nonumber\\
\cc{b_1}{e_5}=\cc{b_2}{e_5}\ ,\ \cc{b_1}{e_6}=\cc{b_2}{e_6}\ , \label{cota}
\end{eqnarray}
where without loss of generality we have assumed that $Q, u^{c}$ 
and $H_u$ arise from 
the sectors $B^1_{0000}=S+b_1$,  $B^2_{0000}=S+b_2$ and  
$B^6_{0000}=S+b_1+b_2$ respectively.
Preserving this coupling, after the breaking of the $SO(10)$ symmetry, 
requires the 
introduction of additional constraints, these are 
\begin{eqnarray}
S^1_{0000}=S^2_{0000}=T^3_{0000}=-1
\label{tpco}
\end{eqnarray}
and
\begin{eqnarray}
& &X^1_{0000}=+1\ ,\ U^{\alpha,1}_{0000}=-1\ ,\ \ U^{\beta,1}_{0000}=-i\ ,\label{tpcoa}\\
& &X^2_{0000}=+1\ ,\ U^{\alpha,2}_{0000}=+1\ ,\ \ U^{\beta,1}_{0000}=+i\ ,\label{tpcob}\\
& &Y^3_{0000}~=+i\ ,\ V^{\beta,3}_{0000}~=\pm1\ ,\ V^{\alpha,3}_{0000}~=-1\ . 
\label{tpcoc}
\end{eqnarray}
Conditions in \eqref{tpco} assert that the states $Q, u^c/H_u$ belong 
to fertile $SO(10)$ 
spinorials/vectorials while conditions \eqref{tpcoa}-\eqref{tpcoc} 
assure the survival of these states after the employment of the 
$\alpha, \beta$ related 
projections according to tables \ref{ptable} and \ref{ptablev}. 
Altogether,
constraints \eqref{tpco}-\eqref{tpcoc} translate to 
\begin{eqnarray}
\cc{b_1}{b_2}=\cc{b_1}{e_5}\, ~~,~~\cc{b_1}{e_6}=-\cc{1}{b_1}\ ,
\nonumber\\
\cc{\alpha}{b_1}=\cc{\beta}{b_1}=-\cc{\alpha}{b_2}=-\cc{\beta}{b_2}=+1\ .
\label{topcriteriab}
\end{eqnarray}
The last two equations together with \eqref{cota} stand for the 
necessary and sufficient 
conditions for the presence of the top quark mass Yukawa coupling in 
the low energy effective field theory of the models under consideration. 

%
\section{\emph{Results}}\label{resu}
In this section we analyse SLM string vacua and classify 
them according 
to their basic phenomenological properties. 
Following the results of Section \ref{analysis2} and restricting to the 
observable spectrum, 
apart from fractional charge exotics, the parameter space
involves 62 GGSO phases taking values $\pm1$ each. A comprehensive scan 
of this space would require examining  $2^{62}\sim 5.\times 10^{18}$ 
configurations.
We note that some redundancy exists in this space of configurations, 
{\it e.g.} with respect to the permutation
symmetry of the $Z_2\times Z_2$ orbifold planes. 
Consequently, some phase configurations may produce identical physical
characteristics. In this paper our objective is to develop the methods 
to extract vacua with specific properties. In particular our focus 
here is with respect to the observable sector, and further development 
of the classification methods with respect to the hidden and exotic sectors
is deferred for future work. We further note that the same redundancy
exists in the classification of the PS and FSU5 free fermionic vacua, 
as they possess a similar $Z_2\times Z_2$ orbifold structure. In comparison
to these cases, the space of SLM phase configurations is vastly increased.
We further comment that one has to ensure that the randomizing 
routine has an appropriately large cycle to reduce the probability that
identical phase configurations are generated.
Despite recent progress in the development of efficient scan algorithms, 
capable of 
scanning up to $10^5$ models per second (see e.g. \cite{fkr}), a full
exploration of this huge parameter space would require thousands of years. 
One strategy for dealing with this problem is to analyse a random sample of the
parameter space and to deduce some conclusions regarding the structure and 
the properties of this class of vacua. An advantage of this method is that
it can be easily adapted to the available computer power and time. Moreover, 
as it is expected that phenomenologically interesting models will in general 
exhibit some degeneracy, a random scan of e.g. $1:10^4$ configurations could 
capture the most important features of these vacua. 
\begin{figure}[h]
\includegraphics[width=\textwidth]{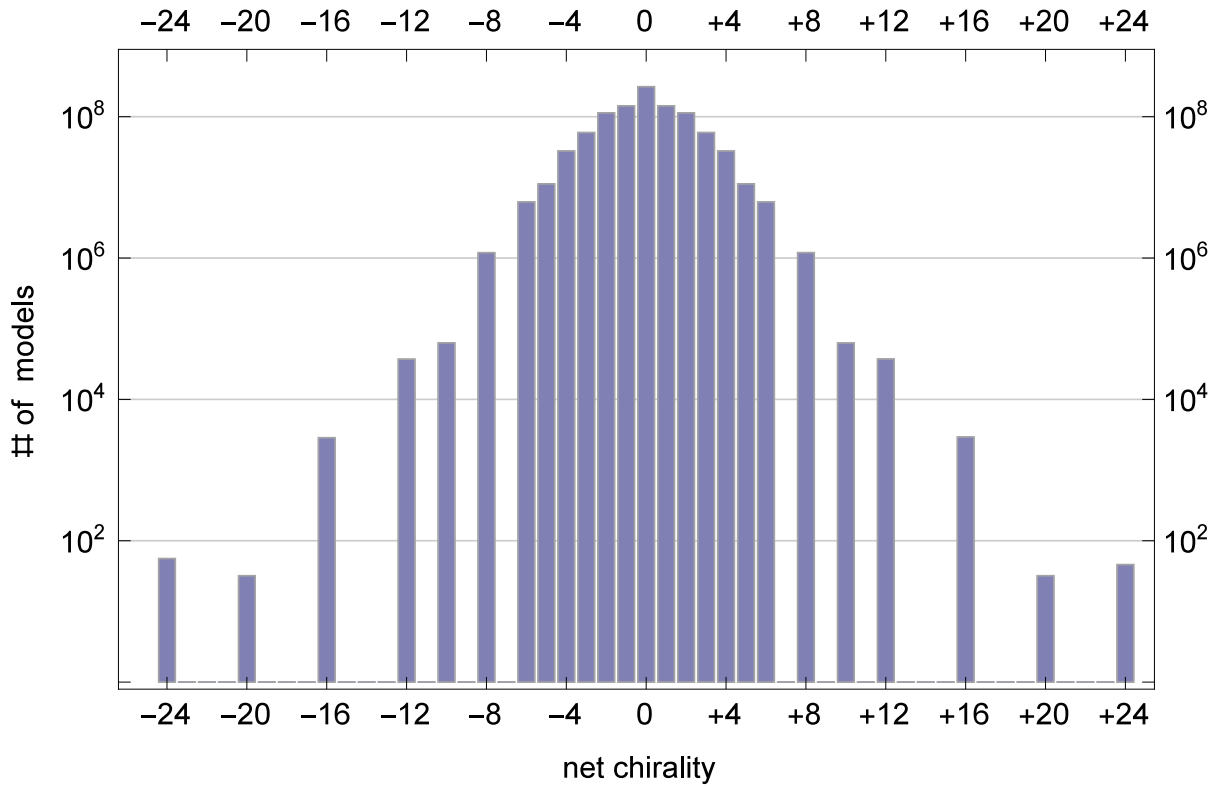}
\caption{\label{chi}
Number of models versus net chirality $N_{16}-N_{\overline{16}}$ in a random 
sample of $10^9$ $SO(10)$ models.}
\end{figure}
In addition, this method has been also successfully applied in the analysis of 
Pati--Salam and flipped $SU(5)$ vacua. However, a straightforward 
implementation in the case of standard model like vacua under consideration turns out to be practically impossible. The reason being that the phenomenologically acceptable models are too rare to be located using fully randomised search. Moreover, 
as will become clearer in the following,
interesting SM vacua are not evenly spaced, but are concentrated 
in small regions of the parameter space, around specific fertile $SO(10)$ cores 
defined in Section  \ref{analysis2}. To demonstrate this we split the parameter space $\Pi$ in a product of two spaces $\Pi=\Pi_1\times{\Pi_2}$. The former $\Pi_1$ comprises GGSO phases that involve the first 12 basis vectors preserving the $SO(10)$ gauge symmetry whilst the latter $\Pi_2$ includes all GGSO phases related to
the $SO(10)$ breaking  vectors $\alpha,\beta$. Following Section 
\ref{analysis2}, $\Pi_1$ includes 44 parameters given in \eqref{parama}, 
whereas $\Pi_2$ consists of the 18 parameters of \eqref{parame}.

\begin{figure}[h]
\includegraphics[width=\textwidth]{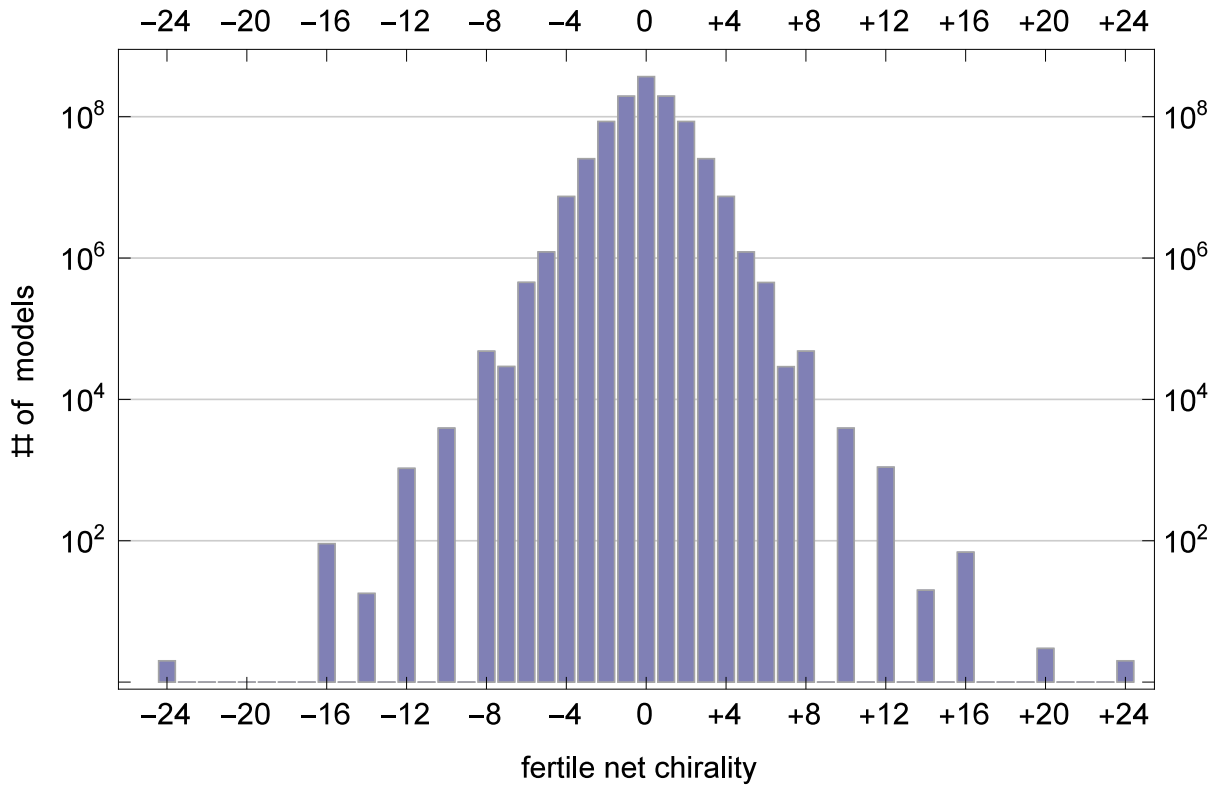}
\caption{\label{fchi}
Number of models versus fertile net chirality $N^f_{16}-N^f_{\overline{16}}$ 
in a random sample of $10^9$ $SO(10)$ configurations.}
\end{figure}

\begin{figure}[h]
\includegraphics[width=\textwidth]{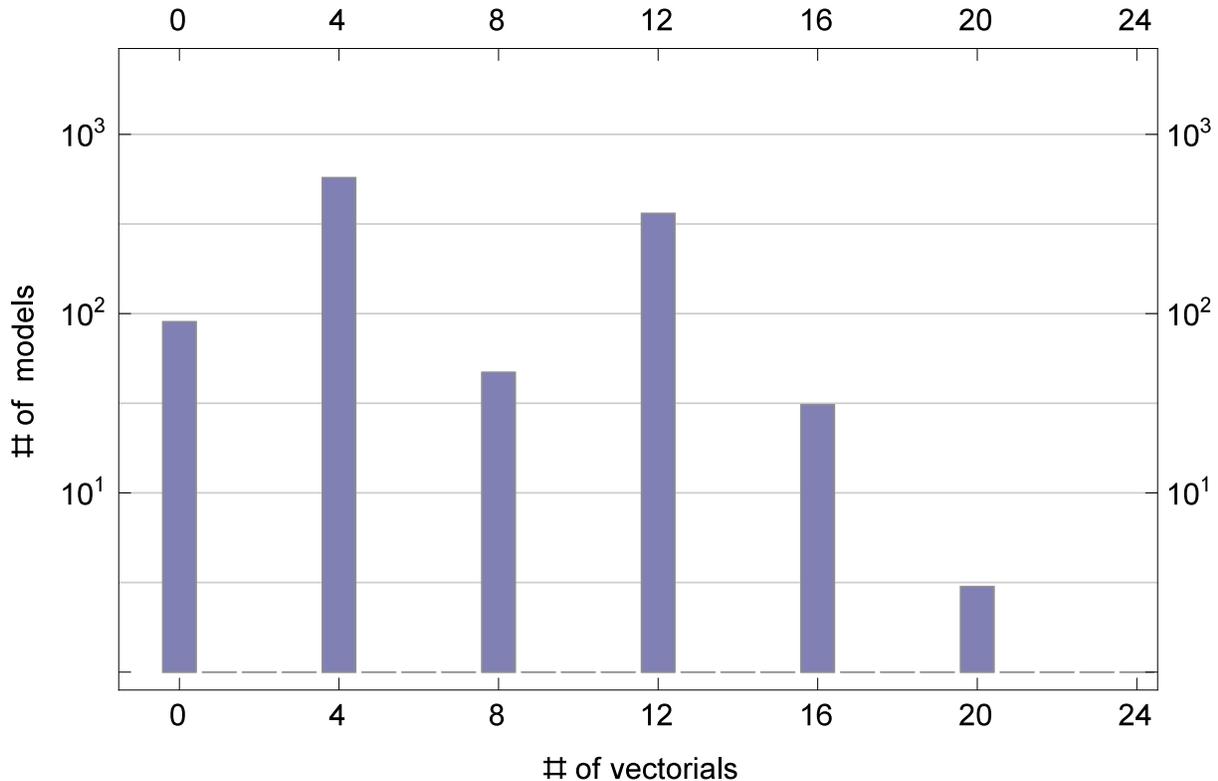}
\caption{\label{vplot}
Number of models with fertile net chirality $N^f_{16}-N^f_{\overline{16}}=12$ 
versus number of twisted fertile vectorial  representations in a random 
sample of $10^9$ $SO(10)$ vacua.}
\end{figure}

Let us focus on the $\Pi_1$ subspace. It comprises $2^{44}\sim 2.\times 10^{13}$ $SO(10)$ configurations. We can apply a random sampling method to study their  basic features.
To this end we have generated a random sample of $10^9$ vacua and calculated the number of spinorial/antispinorial and vectorial representations for each model 
using equations \eqref{spinor}, \eqref{aspinor} and \eqref{vector}. This sampling is quite dense as it comprises approximately one in $10^4$ models of this subspace.
The results for the number of models as a function of the net chirality $N_{16}-N_{\overline{16}}$ are depicted in Figure \ref{chi}. We recover the usual bell shape distribution 
of $SO(10)$ vacua \cite{fkr}.  However, at this point one has to take into account an additional constraint. 
 As explained in Section \ref{analysis2}, when considering the $\beta$ related projections some spinorials 
are entirely projected out and do not give rise to offspring standard model states.
However, these fertile spinorials can be traced back in the $\Pi_1$ parameter space. Thus, the effective net chirality is that of the fertile spinorials $N^f_{16}-N^f_{\overline{16}}$ as defined in \eqref{nss}, \eqref{nsbs}.
We have performed a similar analysis in our random $SO(10)$ model 
sample  and plotted the number of models versus the fertile net chirality in  Figure \ref{fchi}.
Moreover, the final net chirality is also affected by the  
truncation of the SM states accommodated in spinorial representations 
due to the $\alpha,\beta$ projections.
 As can be seen from table \ref{ptable} for fixed values of these projections 
each spinorial is split into four parts out of which only one survives. 
As a result we need at least $4\times3=12$ generations at this level in 
order end up with three generations at the SM level. Consequently, only  
vacua with fertile net chirality 12 can give rise to three
generation models after the application of the $SO(10)$  breaking projections.
Another important phenomenological requirement is the existence of
Higgs doublets in the low energy effective theory spectrum. At least one
massless pair is needed in the minimal supersymmetric scenario. 
Appropriate Higgs doublets are accommodated into $SO(10)$ vectorials 
that arise both from the twisted and the untwisted sectors. However, 
it can be shown that in the class of models under consideration the 
$\alpha$ GGSO projections eliminate all untwisted doublets \cite{dts}. 
Hence, we have to look for the necessary SM Higgs doublets among the 
twisted sector
$SO(10)$ vectorials.
Using similar arguments, as in the case of spinorials, we conclude that
the  number of vectorials that satisfy the GGSO projections related to 
$\Pi_1$ and give rise to Higgs doublets is effectively reduced due to 
two reasons. 
First, some of them become inactive as they do not abide by the  
fertility condition \eqref{fcon}. Second, as vectorials are also 
subject to truncation due to the $\alpha,\beta$ projections
they can give rise to additional triplets instead of doublets.
A look at table \ref{ptablev} is enough to convince us that we need at least
two fertile vectorials at the $SO(10)$ level in order to produce the 
required Higgs doublet pair at the SM level.
A plot of the number of models in our sample with 
$N_{16}-N_{\overline{16}}=12$ versus
the number of $SO(10)$ vectorials is presented in Figure \ref{vplot}.
As seen from the figure a number of models, namely  those with $N^{10}_f=0$, 
in Figure \ref{vplot} fail to comply with this requirement and are thus excluded.
Moreover, there are no models with $N^f_{10}=2$.
Altogether, it turns out that approximately one in a million $SO(10)$ 
configurations in $\Pi_1$ give rise to phenomenologically acceptable offspring SM spectra.

Let us now turn to the $\Pi_2$ subspace. It contains
18 parameters thus it amounts to $2^{18}\sim 2.6\times 10^5$ distinct 
coefficient choices. A preliminary computer search shows that when combined with a
legitimate $\Pi_1$ model they give rise to $2.2\times 10^4$ acceptable 
SM vacua on the average. 
That is one in ten configurations. Altogether
the abundance of acceptable vacua is 
$2.2\times 10^4\times10^3:2.6\times 10^5\times 10^{9}\sim 1:10^{7}$. 
Collecting a reasonable set of say $10^6$ SMs would require examining a sample
of $10^{13}$ configurations. The problem becomes more difficult in practice
as the distribution of acceptable vacua is not homogeneous.  
To resolve this issue we introduce a new search strategy
consisting of a random scan in the parameter space $\Pi_1$ 
combined with a  comprehensive scan of $\Pi_2$. More particularly,
we  first perform a random search in $\Pi_1$ for $SO(10)$ models 
that satisfy the aforementioned constraints
\begin{eqnarray}
N^f_{16}-N^f_{\overline{16}}=12 \ ,\ N^f_{10}\ge 2\ 
\end{eqnarray}
and collect the associated  matching model data.
Afterwords, for each of the assembled configurations we perform a 
comprehensive scan of the parameter space $\Pi_2$ and classify the resulting acceptable model spectra according to their main phenomenological properties. 
At this step, a model is considered as acceptable 
if it satisfies the minimal set of phenomenological criteria  
of Eqs. \eqref{famco}, \eqref{damco}, \eqref{hamco}.
This method turned out to be very efficient. 
A random scan of $10^9$ $SO(10)$ configurations in $\Pi_1$ took 
approximately 8  hours in a computer equipped with Intel i7 CPU 
(4 cores) running at 2.93GHz and 12 GB of RAM and produced 
1011 matching $SO(10)$ models. Then a full scan of the $\Pi_2$ 
parameter space required around 5 additional hours and yielded 
approximately $2.2\times 10^7$ acceptable models.
The main characteristics of these models together with their 
multiplicities
are summarised in table \ref{tresa}. For each different model 
we list $n(Q), n(L), n(d^c)=n(\nu^c), 
n(u^c)=n(e^c)$ i.e. the multiplicities of the associated standard 
model fields together with 
$n(\overline{Q}), n(\overline{L}),
n(\overline{d^c})=n(\overline{\nu^c}),
n(\overline{u^c})=n(\overline{e^c})$ the multiplicities of potential 
fields in conjugate representations 
arising from $SO(10)$ spinorials/antispinorials
as well
as the numbers of Higgs doublet pairs 
$n(H_u)=n(H_d)$ and additional  triplet pairs
 $n({d^c}')=n(\overline{{d^c}'})$ arising from  $SO(10)$ vectorials.

Some comments are in order here concerning the model multiplicity 
in table \ref{tresa}. First, a part of this degeneracy is due to permutation 
symmetry. More particularly, as the basis vectors $v_1,\dots v_{14}$ 
treat the three orbifold planes symmetrically it is expected that 
for every model
with a certain distribution of states in the three twisted planes, 
say $\left(\Xi_1,\Xi_2,\Xi_3\right)$ with $\Xi_I$ the subset of states 
in the I-the twisted plane, there exist 
equivalent models where two of the three subsets are interchanged e.g 
$\left(\Xi_2,\Xi_1,\Xi_3\right)$. We will explain below, in the discussion  
concerning the top quark Yukawa coupling how this degeneracy can be lifted.
Second, in the computation of model multiplicity we have ignored all 
information regarding exotic/fractional charge states, hidden sector states etc. 
Thus, models considered as equivalent in table \ref{tresa} could differ 
substantially with respect to the hidden sector and/or fractional/exotic state
spectra. Third, even in the case where two models have identical spectra they could differ
substantially at the level of interactions. 

\newcolumntype{R}[1]{>{\raggedleft\arraybackslash}p{#1}}
\newcolumntype{C}[1]{>{\centering\arraybackslash}p{#1}}
\begin{table}[!h]
\centering
 \resizebox{15cm}{!}{%
\begin{tabular}{|R{1cm}*{10}{|C{1cm}}|R{2.0cm}|}
\hline
&$n(Q)$&$n(L)$&$n(d^c)$ $n(\nu^c)$
&$n(u^c)$ $n(e^c)$
&$n(\overline{Q})$&$n(\overline{L})$
&$n(\overline{d^c})$ $n(\overline{\nu^c})$
&$n(\overline{u^c})$ $n(\overline{e^c})$&$n(H_u)$ $n(H_d)$&
 $n({d^c}')$  $n(\overline{{d^c}'})$&
Multiplicity\\
\hline
1&3&3&3&3&0&0&0&0&1&1&10748928\\
2&3&3&3&3&0&0&0&0&3&3&7635968\\
3&3&3&3&3&0&0&0&0&2&0&1697792\\
4&3&3&3&3&0&0&0&0&2&2&669696\\
5&3&3&3&3&0&0&0&0&1&5&298496\\
6&3&3&3&3&0&0&0&0&5&1&298496\\
7&4&4&4&4&1&1&1&1&1&1&49152\\
8&3&3&3&3&0&0&0&0&2&4&34816\\
9&3&3&3&3&0&0&0&0&4&2&34816\\
10&3&3&3&3&0&0&0&0&1&3&28672\\
11&3&3&3&3&0&0&0&0&3&1&28672\\
12&3&3&4&4&0&0&1&1&2&0&24576\\
13&4&4&3&3&1&1&0&0&2&0&24576\\
14&4&4&3&3&1&1&0&0&3&5&16640\\
15&3&3&4&4&0&0&1&1&5&3&16640\\
16&3&3&4&4&0&0&1&1&3&5&16640\\
17&4&4&3&3&1&1&0&0&5&3&16640\\
18&3&4&4&3&0&1&1&0&4&4&16384\\
19&4&3&3&4&1&0&0&1&4&4&16384\\
20&3&4&3&4&0&1&0&1&4&4&16384\\
21&4&3&4&3&1&0&1&0&4&4&16384\\
22&3&4&4&3&0&1&1&0&2&2&12288\\
23&4&3&3&4&1&0&0&1&2&2&12288\\
24&3&4&3&4&0&1&0&1&2&2&12288\\
25&4&3&4&3&1&0&1&0&2&2&12288\\
26&3&3&4&4&0&0&1&1&1&1&12288\\
27&4&4&3&3&1&1&0&0&1&1&12288\\
28&3&3&3&3&0&0&0&0&4&0&9216\\
29&4&4&4&4&1&1&1&1&2&0&8192\\
30&4&4&3&3&1&1&0&0&3&1&7680\\
31&3&3&4&4&0&0&1&1&3&1&7680\\
32&4&4&3&3&1&1&0&0&1&3&6144\\
33&3&3&4&4&0&0&1&1&1&3&6144\\
34&3&4&4&3&0&1&1&0&1&1&6144\\
35&4&3&3&4&1&0&0&1&1&1&6144\\
36&3&4&3&4&0&1&0&1&1&1&6144\\
37&4&3&4&3&1&0&1&0&1&1&6144\\
38&3&3&4&4&0&0&1&1&1&7&2816\\
39&4&4&3&3&1&1&0&0&7&1&2816\\
40&3&3&4&4&0&0&1&1&7&1&2816\\
41&4&4&3&3&1&1&0&0&1&7&2816\\
42&3&5&3&3&0&2&0&0&1&3&1536\\
43&5&3&3&3&2&0&0&0&1&3&1536\\
44&3&3&3&5&0&0&0&2&3&1&1536\\
45&3&3&5&3&0&0&2&0&3&1&1536\\
46&3&5&3&3&0&2&0&0&3&1&1536\\
47&5&3&3&3&2&0&0&0&3&1&1536\\
48&3&3&3&5&0&0&0&2&1&3&1536\\
49&3&3&5&3&0&0&2&0&1&3&1536\\
\hline
\end{tabular}}
\caption{\label{tresa}
Distinct standard like models with respect to the  phenomenological 
characteristics under consideration and their multiplicities. 
The models are derived utilising a search over a random sample  of $10^9$  configurations  
in the  $SO(10)$  preserving parameter subspace $\Pi_1$ combined  with
a comprehensive scan
 in the $SO(10)$ breaking parameter subspace $\Pi_2$.}
\end{table}

Another phenomenological  characteristic of particular interest is the
existence of a Yukawa coupling providing mass to the heaviest quark, 
namely the top quark.  
The conditions ensuring the presence of such coupling at the tri-level low energy effective 
superpotential  have been derived in Section \ref{tmy}. They are expressed in terms of 
GGSO phase relations \eqref{cota},\eqref{topcriteriab}. As a result, their implementation  
is straightforward, it suffices to match the
standard like models derived above towards the criteria \eqref{cota},\eqref{topcriteriab}.
The results of this analysis are shown in table \ref{tresa} where we list the main 
characteristics and multiplicities of
distinct standard like models possessing a top quark Yukawa coupling. It turns out that 
almost half of the different models in table \ref{tresa}
are endowed with a top quark mass potential term.
Caution must be taken in comparing the model multiplicities of tables \ref{tresa} 
and \ref{tresatp}. In the implementation of the top quark mass constraints we 
have made certain assumptions about the origin of the states involved ($Q^3, {d^c}^3, H_u$). 
Without loss of generality these assumptions lift some degeneracy 
of the spectra related to twisted plane permutation symmetries. Consequently,
multiplicities in the last column of table \ref{tresatp} have to be 
raised by an extra factor  when compared to those of the 
last column of table \ref{tresa}. 
This factor amounts e.g for assigning $Q^3$ state to $B^1_{0000}$ sector,  ${u^c}^3$ to $B^2_{0000}$ etc.

\begin{table}[!h]
\centering
 \resizebox{15cm}{!}{%
\begin{tabular}{|R{1cm}*{10}{|C{1cm}}|R{2.0cm}|}
\hline
&$n(Q)$&$n(L)$&$n(d^c)$ $n(\nu^c)$
&$n(u^c)$ $n(e^c)$
&$n(\overline{Q})$&$n(\overline{L})$
&$n(\overline{d^c})$ $n(\overline{\nu^c})$
&$n(\overline{u^c})$ $n(\overline{e^c})$&$n(H_u)$ $n(H_d)$&
 $n({d^c}')$  $n(\overline{{d^c}'})$&
Multiplicity\\
\hline
1&3&3&3&3&0&0&0&0&1&1&27264\\
2&3&3&3&3&0&0&0&0&3&3&16896\\
3&3&3&3&3&0&0&0&0&2&0&7296\\
4&3&3&3&3&0&0&0&0&2&2&2304\\
5&3&3&3&3&0&0&0&0&5&1&1536\\
6&4&3&3&4&1&0&0&1&2&2&768\\
7&3&4&3&4&0&1&0&1&2&2&768\\
8&3&3&3&3&0&0&0&0&1&5&640\\
9&4&4&3&3&1&1&0&0&5&3&512\\
10&3&3&4&4&0&0&1&1&3&1&384\\
11&3&3&4&4&0&0&1&1&1&3&384\\
12&3&3&3&3&0&0&0&0&3&1&256\\
13&4&4&3&3&1&1&0&0&3&5&256\\
14&3&3&4&4&0&0&1&1&7&1&192\\
15&4&4&3&3&1&1&0&0&3&1&192\\
16&3&3&3&5&0&0&0&2&3&1&192\\
17&3&3&3&5&0&0&0&2&1&3&192\\
18&3&3&3&3&0&0&0&0&1&3&128\\
19&3&4&3&4&0&1&0&1&4&4&128\\
20&3&4&4&3&0&1&1&0&4&4&128\\
21&4&3&4&3&1&0&1&0&4&4&128\\
22&4&3&3&4&1&0&0&1&4&4&128\\
23&3&3&4&4&0&0&1&1&5&3&64\\
24&3&3&4&4&0&0&1&1&3&5&64\\
25&3&3&4&4&0&0&1&1&1&7&64\\
\hline
\end{tabular}}
\caption{\label{tresatp}
Main phenomenological features and multiplicities of distinct standard--like models 
endowed with a top quark mass Yukawa coupling.
The models are derived utilising a search over a random sample of $10^9$  configurations  in the  $SO(10)$  preserving parameter subspace $\Pi_1$ combined  with
a full scan  in the $SO(10)$ breaking parameter subspace $\Pi_2$. Assumptions have been made with regard to the sectors
producing the associated states (see text for details).
}
\end{table}

As seen from tables \ref{tresa}  three generation SLM vacua display a variety of spectra including : (a) Models without additional twisted triplets, 
as model no 3 in the table. Although, untwisted triplets are not projected out they usually become superheavy through coupling with untwisted sector singlets that acquire vevs.
Hence, these models deserve further study in conjunction with the issue of proton decay. (b) Models with additional vector-like standard model states, including $Q-\bar{Q}$ pairs. 
The presence of these states could raise the SM coupling unification scale to energies close to the string scale. (c) Models with  $\nu^c-\bar{\nu}^c$ pairs. These can play the
role of heavy Higgs that break the additional abelian symmetries giving rise to the standard hyperchange $U(1)_Y$ symmetry. This is a new feature that leads to a new class of 
SLMs that have not been studied previously. Interestingly enough, all above classes of models appear also in table \ref{tresatp}, 
that is they possess a candidate top quark mass Yukawa coupling.
We will study an exemplary model displaying some of the 
above characteristics in the next section.

%
%

\section{\emph{An Exemplary Model}}\label{example}

In this section we use our computerised trawling algorithm to 
extract and discuss one specific model in some detail. The entire
spectrum of the model is derived and presented. The string vacuum
contains three chiral 16 of $SO(10)$ decomposed under the 
$SU(3)\times SU(2)\times U(1)^2$ subgroup, plus the heavy and light
Higgs representations required for realistic symmetry breaking and 
fermion mass generation. Distinctly from previous free fermionic 
SLM constructions, the heavy Higgs states in this model are 
obtained from standard $SO(10)$ representations. The string derived 
model contains an additional pair of vector--like $Q$ and ${\bar Q}$
states that can be used to mitigate the GUT versus heterotic--string
gauge coupling unification problem.
The string model is generated by the set of basis vectors given in 
eqs. (\ref{basis},\ref{alphav}, \ref{betav}),
and by the set of GGSO phases given in eq. (\ref{BigMatrix}). 

\beq \label{BigMatrix}  (v_i|v_j)\ \ =\ \ \bordermatrix{
      & ~1& S&e_1&e_2&e_3&e_4&e_5&e_6&b_1&b_2&z_1&z_2&\alpha&\beta\cr
 1    & ~1& 1& 0& 0& 0& 0& 0& 0& 1& 1& 0& 0& 0& {1\over2}\cr
S     & ~1& 1& 1& 1& 1& 1& 1& 1& 1& 1& 1& 1& 1& 1\cr
e_1   & ~0& 1& 1& 0& 0& 0& 0& 0& 0& 0& 0& 0& 1& 0\cr
e_2   & ~0& 1& 0& 1& 1& 0& 0& 0& 1& 0& 0& 0& 0& 0\cr
e_3   & ~0& 1& 0& 1& 1& 1& 0& 0& 0& 0& 0& 0& 0& 0\cr
e_4   & ~0& 1& 0& 0& 1& 1& 0& 0& 0& 0& 1& 0& 0& 0\cr
e_5   & ~0& 1& 0& 0& 0& 0& 1& 0& 0& 0& 0& 0& 1& 0\cr
e_6   & ~0& 1& 0& 0& 0& 0& 0& 1& 0& 0& 0& 0& 0& 1\cr
b_1   & ~1& 0& 0& 1& 0& 0& 0& 0& 1& 0& 1& 0& 0& {1\over2}\cr
b_2   & ~1& 0& 0& 0& 0& 0& 0& 0& 0& 1& 0& 0& 0&-{1\over2}~~\cr
z_1   & ~0& 1& 0& 0& 0& 1& 0& 0& 1& 0& 0& 1& 0& {1\over2}\cr
z_2   &~ 0& 1& 0& 0& 0& 0& 0& 0& 0& 0& 1& 0& 0& 1\cr
\alpha& ~0& 1& 1& 0& 0& 0& 1& 0& 1& 1& 1& 0& 0& 1\cr
\beta & ~1& 1& 0& 0& 0& 0& 0& 1& 0& 1& 0& 0& 0& {1\over2} \cr
  }
\eeq
where we used the notation
$\cc{v_i}{v_j} = e^{i\pi (v_i|v_j)}$.
The spacetime vector bosons in the model are obtained from 
three sectors: the Neveu--Schwarz (NS) sector; the $z_2$--sector; 
and the $x+2\beta$--sector\footnote{Below we use the definition
$z_3=x+2\beta$.}.
The vector bosons from the NS--sector
generate the observable and hidden sector symmetries given
in eqs. (\ref{ogaugesym}) and (\ref{hgaugesym}). The $z_2$--sector
enhances the $SU(2)_{h2}\times SU(2)_{h3}\times U(1)_{SU(4)_h}$ to 
an hidden $SU(4)_h$ gauge symmetry, whereas the vector boson
states from the $z_3$--sector enhance the 
$SU(2)_{h1}\times SU(2)_{h4}$, together with the real fermion
$\bar\omega_2$ to $SO(5)$. 
The $U(1)$ combinations are: 
\beqn
U(1)_{SU(4)}        & = & U_{h4}+U_{h2}+U_{h3}~, \\
U(1)_{5^\prime} ~~~~& = & U_{h4}-U_{h2}-U_{h3}~, \\
U(1)_{6^\prime} ~~~~& = & ~~~~~~~~U_{h2}-U_{h3}~. 
\eeqn
We emphasize that the non--simply laced $SO(5)$ symmetry is 
generated due to the fact that the states from the $z_3$--sector,
that enhance the untwisted gauge symmetry, are obtained by 
acting on the vacuum with the real fermion oscillator $\bar\omega_2$. 
One generator in the Cartan sub--algebra 
is projected out and the roots are not charged under the 
associated broken $U(1)$ symmetry. Consequently, the roots 
obtained from the $z_3$--sector have length 1 and the resulting
group is non--simply laced. 
Extensive investigations of using real fermions in similar constructions
are discussed in ref. \cite{johndienes}. 
The full massless matter spectrum of the model is displayed in 
tables \ref{tablea}, \ref{tableb},  \ref{tableslmexotics}, 
\ref{tablePSe}, \ref{tablefsu5e}, \ref{tableso10singlets}.
The model possess $N=1$ spacetime supersymmetry and therefore
all the states shown in the table are in super--multiplets. 
Table \ref{tablea} shows the untwisted matter states that are 
charged under the observable gauge group. A single untwisted 
state, $V_{33}$,  which is charged under the hidden sector 
gauge group is shown in table \ref{tableso10singlets}.
Table \ref{tableb} shows the observable matter states. 
The states in table \ref{tableb} are charged only under the 
observable gauge symmetry in eq. (\ref{ogaugesym}) but
not under the hidden gauge symmetry in eq. (\ref{hgaugesym}). 
As seen from table \ref{tableb} the model contains three
chiral generations and the required heavy and light Higgs 
states for $U(1)_{Z^\prime}$ and electroweak symmetry breaking. 
The observable spectrum of this model
exhibits several novel features compared to the earlier
SLM free fermionic constructions \cite{slm}.
The model contains the state ${\bar N}_1$, which together with 
a combination of the $N_i,~i=1,\cdots,4$ can be used to break the 
$U(1)_{Z^\prime}$ symmetry along flat directions. 
This should be contrasted with the earlier SLM models in which such
a state was absent. Those models therefore necessarily utilised
exotic states that carry fractional $U(1)_{Z^\prime}$ charge.
Breaking the $U(1)_{Z^\prime}$ gauge symmetry with 
states that carry standard GUT charges leaves a remnant 
local discrete symmetry \cite{lds} that protects the 
exotic states from decaying into the Standard Model states. 
In this case the exotic states provide viable dark matter 
candidates \cite{ccr}. However, in the absence of states
with standard GUT charges, exotically charged states are 
utilised \cite{slm}, which does not leave a remnant 
discrete symmetry. The dark matter scenario of ref. 
\cite{ccr} was recently realised in \cite{dfmr} with states
that are exotic with respect to $E_6$, but are singlets 
under the $SO(10)$ gauge group, {\it i.e.} these states 
are neutral under the $U(1)_{Z^\prime}$ symmetry of eq. (\ref{u1zprime}).
The model presented 
here therefore provides examples of viable dark matter candidates 
that are Standard Model singlets and are charged under this
$U(1)_{Z^\prime}$ combination. These states are shown in 
table \ref{tableslmexotics}. The second novel property of this model compared
to the earlier construction of \cite{slm} is the 
additional pair of $Q$ and ${\bar Q}$, that may play a role 
in resolving the discrepancy between the GUT and heterotic string 
unification scales \cite{gutu, gcu}. 

The states displayed in table \ref{tableso10singlets}
are singlets of $SO(10)$ and hence neutral under the 
Standard Model subgroup. They are charged with respect to the
observable and hidden $U(1)$ gauge symmetries and may transform 
in non--Abelian representations of the hidden $SU(4)\times SO(5)$ 
gauge symmetry. 
The last state appearing in table \ref{tableso10singlets}, $V_{33}$, 
is obtained from the untwisted sector, whereas all other states
are obtained from the twisted sectors. The untwisted state 
$V_{33}$ arises due to the gauge symmetry enhancement
from the $z_3$--sector. 

The states displayed in tables \ref{tableslmexotics}, \ref{tablePSe} and
\ref{tablefsu5e} are exotic states that arise due to the Wilson
line breaking of the $SO(10)$ GUT symmetry. As discussed in section 
\ref{analysis4} these states are classified according to the 
$SO(10)$ subgroup that is left unbroken in the sectors from which they
arise. The states in tables \ref{tablePSe} and \ref{tablefsu5e}
leave unbroken the $SO(6)\times SO(4)$ and $SU(5)\times U(1)$ 
subgroups, respectively, and therefore also arise in the 
free fermionic Pati--Salam \cite{alr, acfkr} and flipped $SU(5)$
\cite{fsu5, gutu, frs, hasan} type models. The states from these
sectors carry fractional electric charge $\pm1/2$, which are 
highly constrained by observations \cite{vhalyo}. 
We note that a proposed resolution
is that all the fractionally charged states transform in non--Ableian 
representations of the hidden sector gauge group and are confined 
into integrally charged states \cite{cryptons}. 
This is similar to the situation with the 
SLM exotics in table \ref{tableslmexotics}, which all transform under the 
hidden $SU(4)$ gauge symmetry. Indeed, that is also the case with the 
fractionally charged states appearing in table \ref{tableslmexotics}. 
However, while this is indeed the case in the flipped $SU(5)$ model 
of ref. \cite{fsu5}, it does not in general hold in the space of 
flipped $SU(5)$ \cite{frs} or Pati--Salam heterotic--string vacua \cite{acfkr}. 
An alternative possibility is that the fractionally charged states
obtain string scale mass from effective mass terms in the superpotential
\cite{fc}. The most compelling possibility, however, is that
fractionally charged states appear as massive states in the string
spectrum, but not at the massless level. Indeed, such Pati--Salam 
models were found in ref. \cite{acfkr, su62, frzprime} and were
dubbed exophobic string vacua. As seen from tables \ref{tablePSe} and
\ref{tablefsu5e} the present model contains a variety
of fractionally charged states.

\begin{table}[!t]
\noindent
{\small
\openup\jot
\begin{tabular}{|l|l|c|c|c|c|}
\hline
sector&field&$SU(3)\times{SU(2)}_L\times{U(1)}_C\times U(1)_L$
&${U(1)}_1$&${U(1)}_2$&${U(1)}_3$\\
\hline
$S$&$D_1$&$(3,\hphantom{+}1,{-1},\hphantom{+} 0)$
                                  &$+1$&$\hphantom{+}0$&$\hphantom{+}0$\\
&$D_2$&$(3,\hphantom{+}1,{-1},\hphantom{+} 0)$&$\hphantom{+}0$
                                               &$+1$&$\hphantom{+}0$\\
&$D_3$&$(3,\hphantom{+}1,{-1},\hphantom{+} 0)$&$\hphantom{+}0$
                                                  &$\hphantom{+}0$&$+1$\\
&$\bar{D}_1$&$({\bar3},\hphantom{+}1,+1,\hphantom{+} 0)$
                                  &$-1$&$\hphantom{+}0$&$\hphantom{+}0$\\
&$\bar{D}_2$&$({\bar3},\hphantom{+}1,{+1},\hphantom{+} 0)$
                                  &$\hphantom{+}0$&$-1$&$\hphantom{+}0$\\
&$\bar{D}_3$&$({\bar3},\hphantom{+}1,{+1},\hphantom{+} 0)$
                                  &$\hphantom{+}0$&$\hphantom{+}0$&$-1$\\
&$\Phi_{12}$&$(1,\hphantom{+}1,\hphantom{+}0,\hphantom{+}0 )$
                                  &$+1$&$-1$&$\hphantom{+}0$\\
&$\bar{\Phi}_{12}$&$(1,\hphantom{+}1,\hphantom{+}0,\hphantom{+}0)$
                                              &$-1$&$+1$&$\hphantom{+}0$\\
&$\Phi_{13}$&$(1,\hphantom{+}1,\hphantom{+}0,\hphantom{+}0 )$
                                               &$+1$&$\hphantom{+}0$&$-1$\\
&$\bar{\Phi}_{13}$&$(1,\hphantom{+}1,\hphantom{+}0,\hphantom{+}0)$
                                               &$-1$&$\hphantom{+}0$&$+1$\\
&$\Phi_{23}$&$(1,\hphantom{+}1,\hphantom{+}0,\hphantom{+}0 )$
                                                &$\hphantom{+}0$&$-1$&$+1$\\
&$\bar{\Phi}_{23}$&$(1,\hphantom{+}1,\hphantom{+}0,\hphantom{+}0)$
                                               &$\hphantom{+}0$&$+1$&$-1$\\
&$\Phi_i,i=1,\dots,5$&$(1,\hphantom{+}1,\hphantom{+}0,\hphantom{+}0)$
                               &$\hphantom{+}0$&$\hphantom{+}0$&$\hphantom{+}0$\\
\hline
\end{tabular}
}
\caption{\label{tablea}\it
Observable untwisted matter spectrum and
$SU(3)_C\times{SU(2)}_L\times{U(1)}_C\times U(1)_L\times{U(1)}^3$ 
quantum numbers. }
\end{table}

\begin{table}[!h]
\noindent
{\small
\openup\jot
\begin{tabular}{|l|l|c|c|c|c|}
\hline
sector&field&$SU(3)\times{SU(2)}_L\times{U(1)}_C\times U(1)_L$
                                  &${U(1)}_1$&${U(1)}_2$&${U(1)}_3$\\
\hline
$S+b_1+e_3+e_4+e_6$&${U}_{1}$&$({\bar3},\hphantom{+}1,-{1/2}, -1)$
                                  &$-{1/2}$&$\hphantom{+}0$&$\hphantom{+}0$\\
$ $&${E}_{1}$&$(1,\hphantom{+}1,{\hphantom{+}{3}/{2}}, +1)$
                                  &$-{1/2}$&$\hphantom{+}0$&$\hphantom{+}0$\\
\hline
$S+b_1+e_3+e_4$&${D}_{1}$&$({\bar3},\hphantom{+}1,{-{1/2}}, +1)$
                                  &$-{1/2}$&$\hphantom{+}0$&$\hphantom{+}0$\\
$ $&${ N}_{1}$&$(1,\hphantom{+}1,{\hphantom{+}{3/2}}, -1)$
                                      &$-{1/2}$&$\hphantom{+}0$&$\hphantom{+}0$\\
\hline
$S+b_1+e_3+e_4+e_5+e_6$&$L_{1}$&$(1,\hphantom{+}2,{-{3/2}},\hphantom{+}0 )$
                                                &$-{1/2}$&$\hphantom{+}0$&$\hphantom{+}0$\\
\hline
$S+b_1+e_3+e_4+e_5$&$Q_{1}$&$(3,\hphantom{+}2,{\hphantom{+}{1/2}},\hphantom{+}0 )$
                                                &$-{1/2}$&$\hphantom{+}0$&$\hphantom{+}0$\\
\hline
$S+b_2+e_1+e_5$&${U}_{2}$&$({\bar3},\hphantom{+}1,{-{1/2}}, -1)$
                                                &$\hphantom{+}0$&$-{1/2}$&$\hphantom{+}0$\\
$ $&${E}_{2}$&$(1,\hphantom{+}1,{\hphantom{+}{3/2}}, +1)$
                                                &$\hphantom{+}0$&$-{1/2}$&$\hphantom{+}0$\\
\hline
$S+b_2+e_1+e_5+e_6$&${D}_{2}$&$({\bar3},\hphantom{+}1,{-{1/2}}, +1)$
                                                &$\hphantom{+}0$&$-{1/2}$&$\hphantom{+}0$\\
$ $&${N}_{2}$&$(1,\hphantom{+}1,{\hphantom{+}{3/2}}, -1)$
                                                &$\hphantom{+}0$&$-{1/2}$&$\hphantom{+}0$\\
\hline
$S+b_2+e_1$&$L_{2}$&$(1,\hphantom{+}2,{-{3/2}},\hphantom{+}0 )$
                                                &$\hphantom{+}0$&$-{1/2}$&$\hphantom{+}0$\\
\hline
$S+b_2+e_1+e_6$&$Q_{2}$&$(3,\hphantom{+}2,{\hphantom{+}{1/2}},\hphantom{+}0 )$
                                                &$\hphantom{+}0$&$-{1/2}$&$\hphantom{+}0$\\
\hline
$S+b_2+e_1+e_5$&${U}_{3}$&$({\bar3},\hphantom{+}1,{-{1/2}}, -1)$
                                                &$\hphantom{+}0$&$-{1/2}$&$\hphantom{+}0$\\
$ $&${E}_{3}$&$(1,\hphantom{+}1,{\hphantom{+}{3/2}}, +1)$
                                                &$\hphantom{+}0$&$-{1/2}$&$\hphantom{+}0$\\
\hline
$S+b_2+e_1+e_5+e_6$&${ D}_{3}$&$({\bar3},\hphantom{+}1,{-{1/2}}, +1)$
                                                &$\hphantom{+}0$&$-{1/2}$&$\hphantom{+}0$\\
$ $&${ N}_{3}$&$(1,\hphantom{+}1,{\hphantom{+}{3/2}}, -1)$
                                                &$\hphantom{+}0$&$-{1/2}$&$\hphantom{+}0$\\
\hline
$S+b_2+e_1$&$L_{3}$&$(1,\hphantom{+}2,{-{3/2}},\hphantom{+}0 )$
                                                &$\hphantom{+}0$&$-{1/2}$&$\hphantom{+}0$\\
\hline
$S+b_2+e_1+e_6$&$Q_{3}$&$(3,\hphantom{+}2,{\hphantom{+}{1/2}},\hphantom{+}0 )$
                                                &$\hphantom{+}0$&$-{1/2}$&$\hphantom{+}0$\\
\hline
$S+b_3+e_1$&${ D}_{4}$&$({\bar 3},\hphantom{+}1,{-{1/2}}, \hphantom{+}1)$
                                                &$\hphantom{+}0$&$\hphantom{+}0$&$-{1/2}$\\
$S+b_3+e_1$&${N}_{4}$&$(1,\hphantom{+}1,{\hphantom{+}{3/2}}, -1)$
                                                &$\hphantom{+}0$&$\hphantom{+}0$&$-{1/2}$\\
\hline
$S+b_3$&$Q_{4}$&$(3,\hphantom{+}2,{\hphantom{+}{1/2}},\hphantom{+}0 )$
                                                &$\hphantom{+}0$&$\hphantom{+}0$&$-{1/2}$\\
\hline
$S+b_3+e_1+e_3$&${\bar D}_{1}$&$({3},\hphantom{+}1,\hphantom{+}{1/2}, -1)$
                                                &$\hphantom{+}0$&$\hphantom{+}0$&$\hphantom{+}{1/2}$\\
$S+b_3+e_1+e_3$&${\bar N}_{1}$&$(1,\hphantom{+}1,{-{3}/{2}}, +1)$
                                                &$\hphantom{+}0$&$\hphantom{+}0$&$\hphantom{+}{1/2}$\\
\hline
$S+b_3+e_3$&${\bar Q}_{1}$&$({\bar3},\hphantom{+}2,{-{1/2}},\hphantom{+}0 )$
                                                &$\hphantom{+}0$&$\hphantom{+}0$&$\hphantom{+}{1/2}$\\
\hline
$S+b_3+x+e_4$&${\cal D}_1$&$({ 3},\hphantom{+}1,~~{-1}, \hphantom{+}0)$
                                          &$-{1/2}$&$-{1/2}$&$\hphantom{+}0$\\
$S+b_3+x+e_4$&$S_1$&$({ 1},\hphantom{+}1,~~\hphantom{+}0, \hphantom{+}0)$
                                          &$\hphantom{+}{1/2}$&$\hphantom{+}{1/2}$&$+1$\\
$S+b_3+x+e_3+e_4$&${\bar {\cal D}}_1$&$({\bar 3},\hphantom{+}1,~~{+1}, \hphantom{+}0)$
                                          &$\hphantom{+}{1/2}$&$\hphantom{+}{1/2}$&$\hphantom{+}0$\\
$S+b_3+x+e_3+e_4$&${\bar S}_1$&$({ 1},\hphantom{+}1,~~\hphantom{+}0, \hphantom{+}0)$
                                          &$-{1/2}$&$-{1/2}$&$-1$\\
$S+b_3+x+e_1+e_4$&${ h}_1$&$({ 1},\hphantom{+}2,~~\hphantom{+}0, -1)$
                                          &$\hphantom{+}{1/2}$&$\hphantom{+}{1/2}$&$\hphantom{+}0$\\
$S+b_3+x+e_3+e_4$&${\bar h}_1$&$({ 1},\hphantom{+}2,~~\hphantom{+}0, +1)$
                                          &$\hphantom{+}{1/2}$&$\hphantom{+}{1/2}$&$\hphantom{+}0$\\
\hline
$S+b_1+x+e_3+e_5+e_6$&$\zeta_1$&$({ 1},\hphantom{+}1,~~\hphantom{+}0, \hphantom{+}0)$
                                                 &$\hphantom{+}0$&$\hphantom{+}{1/2}$&$-{1/2}$\\
\hline
$S+b_1+x+e_5+e_6$&${\cal D}_2$&$({ 1},\hphantom{+}1,~~-1, \hphantom{+}0)$
                                          &$\hphantom{+}0$&$\hphantom{+}{1/2}$&$\hphantom{+}{1/2}$\\
$S+b_1+x+e_5+e_6$&$\zeta_a,~a=2,3$&$({ 1},\hphantom{+}1,~~~\hphantom{+}0, \hphantom{+}0)$
                                                 &$\hphantom{+}0$&$\hphantom{+}{1/2}$&$-{1/2}$\\
$S+b_1+x+e_5+e_6$&${\bar\zeta}_a,~a=2,3$&$({ 1},\hphantom{+}1,~~~\hphantom{+}0, \hphantom{+}0)$
                                                 &$\hphantom{+}0$&$-{1/2}$&$\hphantom{+}{1/2}$\\
$S+b_1+x+e_5+e_6$&$\phi_1$&$({ 1},\hphantom{+}1,~~~\hphantom{+}0, \hphantom{+}0)$
                                                 &$+1$&$-{1/2}$&$-{1/2}$\\
$S+b_1+x+e_5$&${\bar {\cal D}}_2$&$({ 1},\hphantom{+}1,~~+1, \hphantom{+}0)$
                                          &$\hphantom{+}0$&$\hphantom{+}{1/2}$&$\hphantom{+}{1/2}$\\
$S+b_1+x+e_5$&${\bar S}_2$&$({ 1},\hphantom{+}1,~~+1, \hphantom{+}0)$
                                          &$-1$&$-{1/2}$&$-{1/2}$\\
\hline
$S+b_1+x+e_6$&${h}_2$&$({ 1},\hphantom{+}2,~~\hphantom{+}0, -1)$
                                          &$\hphantom{+}0$&$\hphantom{+}{1/2}$&$\hphantom{+}{1/2}$\\
$S+b_1+x$&${\bar h}_2$&$({ 1},\hphantom{+}2,~~\hphantom{+}0, +1)$
                                          &$\hphantom{+}0$&$\hphantom{+}{1/2}$&$\hphantom{+}{1/2}$\\

\hline
\end{tabular}
}
\caption{\label{tableb}\it
Twisted matter spectrum (observable sector)  and
$SU(3)_C\times{SU(2)}_L\times{U(1)}_C\times U(1)_L\times{U(1)}^3$ quantum numbers. }
\end{table}

\newpage
\normalsize
\afterpage{%
    \clearpage
    \begin{landscape}
        \centering 
\begin{table}
\noindent
{\small
\openup\jot
\begin{tabular}{|l|l|c|c|c|c|
                              c|c|c|
                                     c|c|c|c|}
\hline
sector&field&$SU(3)\times{SU(2)}$ &${U(1)}_C$ &$U(1)_L$
                                     &${U(1)}_1$&${U(1)}_2$&${U(1)}_3$
                                     &$SU(4)\times SO(5)$ 
                                 & $U(1)_4$& $U(1)_{5^\prime}$& $U(1)_{6^\prime}$\\
\hline
$S+b_2+\alpha\pm\beta+(z_2)$
                         &${\cal S}_{1}$               
                         &$({1},\hphantom{+}1)$
                         &${-{3/4}}$ &$\hphantom{+}{{1/2}}$
                         &$\hphantom{+}{1/4}$&$-{1/4}$&$\hphantom{+}{1/4}$
                         &$(4, 1) $
                         &$-{1/2}$ &$ \hphantom{+}0$ & $\hphantom{+}{1/2}$\\
\hline
$S+b_2+\alpha\pm\beta+(z_2)+e_1+e_5$
                         &${\bar {\cal S}}_{1}$          
                         &$({1},\hphantom{+}1)$
                         &$\hphantom{+}{3/4}$ &$-{1/2}$
                         &$-{1/4}$&$\hphantom{+}{1/4}$&$-{1/4}$
                         &$({\bar4}, 1) $
                         &$\hphantom{+}{1/2}$ &$ \hphantom{+}0$ & $-{1/2}$\\
\hline
$S+b_2+x+\alpha\pm\beta+(z_2)+e_1$
                         &${\cal S}_{2}$                   
                         &$({1},\hphantom{+}1)$
                         &$-{3/4}$ &$\hphantom{+}{1/2}$
                         &$\hphantom{+}{1/4}$&$-{1/4}$&$\hphantom{+}{1/4}$
                         &$({4}, 1) $
                         &$\hphantom{+}{1/2}$ &$ \hphantom{+}0$ & $-{1/2}$\\
\hline
$S+b_2+x+\alpha\pm\beta+(z_2)+e_5     $
                        &${\bar {\cal S}}_{2}$              
                        &$({1},\hphantom{+}1)$
                        &$\hphantom{+}{3/4}$ &$-{1/2}$
                        &$-{1/4}$&$\hphantom{+}{1/4}$&$-{1/4}$
                        &$({\bar4}, 1) $
                        &$-{1/2}$ &$\hphantom{+}0$ & $\hphantom{+}{1/2}$\\%
\hline
\end{tabular}
}
\caption{\label{tableslmexotics}\it
Twisted standard--like exotic matter states and their charges under
the observable 
$SU(3)_C\times{SU(2)}_L\times{U(1)}_C\times U(1)_L\times{U(1)}^3$ 
and the hidden
${SU(4)}\times{SO(5)}\times U(1)^3$ 
gauge symmetries. These exotic states carry standard charges under the 
Standard Model gauge symmetry, but carry non--$SO(10)$ charges under the 
$U(1)_{Z^\prime}$ combination, which is embedded in $SO(10)$. 
}
\end{table}
    \end{landscape}
    \clearpage
}
\newpage
\normalsize
\afterpage{%
    \clearpage
    \begin{landscape}
        \centering 
\begin{table}
\noindent
{\small
\openup\jot
\begin{tabular}{|l|l|c|c|c|c|
                              c|c|c|
                                     c|c|c|c|}
\hline
sector&field&$SU(3)\times{SU(2)}$ &${U(1)}_C$ &$U(1)_L$
                                   &${U(1)}_1$&${U(1)}_2$&${U(1)}_3$
                                   &$SU(4)\times SO(5)$ 
                                   & $U(1)_4$& $U(1)_{5^\prime}$& $U(1)_{6^\prime}$\\
\hline
$S+b_2+\alpha+2\beta+z_2+e_5$
                        &${\cal P}_{1}$                  
                        &$({1},\hphantom{+}2)$
                        &$\hphantom{+}0$ &$\hphantom{+}0$
                        &$-{1/2}$&$\hphantom{+}0$&$-{1/2}$
                        &$(1, 1) $
                        &$\hphantom{+}0$ &$\hphantom{+}0$ & $+1$\\
\hline
$S+b_2+\alpha+2\beta+z_2+e_5+e_6$
                        &${\cal P}_{2}$                      
                        &$({1},\hphantom{+}2)$
                        &$\hphantom{+}0$ &$\hphantom{+}0$
                        &$-{1/2}$&$\hphantom{+}0$&$-{1/2}$
                        &$(1, 1) $
                        &$\hphantom{+}0$ &$\hphantom{+}0$ & $-1$\\
\hline
$S+b_2+\alpha+2\beta+z_2+e_1$
                         &${\cal P}_{3}$                     
                         &$({1},\hphantom{+}2)$
                         &$\hphantom{+}0$ &$\hphantom{+}0$
                       &$\hphantom{+}{1/2}$&$\hphantom{+}0$&$\hphantom{+}{1/2}$
                         &$(1, 1) $
                      &$\hphantom{+}0$&$\hphantom{+}0$&$-1$\\
\hline
$S+b_2+\alpha+2\beta+z_2+e_1+e_6$
                         &${\cal P}_{4}$                    
                         &$({1},\hphantom{+}2)$
                         &$\hphantom{+}0$ &$\hphantom{+}0$
                       &$\hphantom{+}{1/2}$&$\hphantom{+}0$&$\hphantom{+}{1/2}$
                         &$(1, 1) $
                      &$\hphantom{+}0$&$\hphantom{+}0$&$+1$\\
\hline
$S+b_2+x+\alpha+e_1+e_5+e_6$
                         &${\cal P}_{5}$                    
                         &$({1},\hphantom{+}2)$
                         &$\hphantom{+}0$ &$\hphantom{+}0$
                       &$-{1/2}$&$\hphantom{+}0$&$-{1/2}$
                         &$(1, 1) $
                      &$-1$&$\hphantom{+}0$&$\hphantom{+}0$\\
\hline
$S+b_2+x+\alpha+e_1+e_5$
                         &${\cal P}_{6}$                     
                         &$({1},\hphantom{+}2)$
                         &$\hphantom{+}0$ &$\hphantom{+}0$
                       &$-{1/2}$&$\hphantom{+}0$&$-{1/2}$
                         &$(1, 1) $
                      &$+1$&$\hphantom{+}0$&$\hphantom{+}0$\\
\hline
$S+b_2+x+\alpha+e_6$
                         &${\cal P}_{7}$                        
                         &$({1},\hphantom{+}2)$
                         &$\hphantom{+}0$ &$\hphantom{+}0$
                 &$\hphantom{+}{1/2}$&$\hphantom{+}0$&$\hphantom{+}{1/2}$
                         &$(1, 1) $
                      &$+1$&$\hphantom{+}0$&$\hphantom{+}0$\\
\hline
$S+b_2+x+\alpha$
                         &${\cal P}_{8}$                          
                         &$({1},\hphantom{+}2)$
                         &$\hphantom{+}0$ &$\hphantom{+}0$
                 &$\hphantom{+}{1/2}$&$\hphantom{+}0$&$\hphantom{+}{1/2}$
                         &$(1, 1) $
                      &$-1$&$\hphantom{+}0$&$\hphantom{+}0$\\
\hline
$S+b_2+\alpha+2\beta+z_2$
                         &${\cal P}_{9}$                      
                         &$({1},\hphantom{+}1)$
                         &$\hphantom{+}0$ &$-1$
                        &$-{1/2}$&$\hphantom{+}0$&$\hphantom{+}{1/2}$
                        &$(1, 1) $
                        &$\hphantom{+}0$ &$\hphantom{+}0$ & $+1$\\
$                      $
                        &${\cal P}_{10}$                      
                        &$({1},\hphantom{+}1)$
                        &$\hphantom{+}0$ &$+1$
                        &$-{1/2}$&$\hphantom{+}0$&$\hphantom{+}{1/2}$
                        &$(1, 1) $
                        &$\hphantom{+}0$ &$\hphantom{+}0$ & $-1$\\
%
\hline
$S+b_2+\alpha+2\beta+z_2+e_6$
                        &${\cal P}_{11}$                  
                        &$({1},\hphantom{+}1)$
                        &$\hphantom{+}0$ &$-1$
                        &$-{1/2}$&$\hphantom{+}0$&$\hphantom{+}{1/2}$
                        &$(1, 1) $
                        &$\hphantom{+}0$ &$\hphantom{+}0$ & $-1$\\
$                      $
                        &${\cal P}_{12}$                      
                        &$({1},\hphantom{+}1)$
                        &$\hphantom{+}0$ &$+1$
                        &$-{1/2}$&$\hphantom{+}0$&$\hphantom{+}{1/2}$
                        &$(1, 1) $
                        &$\hphantom{+}0$ &$\hphantom{+}0$ & $+1$\\
\hline
$S+b_2+\alpha+2\beta+z_2+e_1+e_5+e_6$
                        &${\cal P}_{13}$                      
                        &$({1},\hphantom{+}1)$
                        &$\hphantom{+}0$ &$-1$
                        &$\hphantom{+}{1/2}$&$\hphantom{+}0$&$-{1/2}$
                        &$(1, 1) $
                        &$\hphantom{+}0$ &$\hphantom{+}0$ & $-1$\\
$                      $
                        &${\cal P }_{14}$                        
                        &$({1},\hphantom{+}1)$
                        &$\hphantom{+}0$ &$+1$
                        &$\hphantom{+}{1/2}$&$\hphantom{+}0$&$-{1/2}$
                        &$(1, 1) $
                        &$\hphantom{+}0$ &$\hphantom{+}0$ & $+1$\\
\hline
$S+b_2+x+\alpha+e_1+e_6$
                         &${\cal P}_{15}$                       
                        &$({1},\hphantom{+}1)$
                         &$\hphantom{+}0$ &$-1$
                       &$-{1/2}$&$\hphantom{+}0$&$\hphantom{+}{1/2}$
                         &$(1, 1) $
                      &$-1$&$\hphantom{+}0$&$\hphantom{+}0$\\
$ $
                         &${\cal P}_{16}$                      
                         &$({1},\hphantom{+}1)$
                         &$\hphantom{+}0$ &$+1$
                       &$-{1/2}$&$\hphantom{+}0$&$\hphantom{+}{1/2}$
                         &$(1, 1) $
                      &$+1$&$\hphantom{+}0$&$\hphantom{+}0$\\
\hline
$S+b_2+x+\alpha+e_1      $
                        &${\cal P}_{17}$                        
                        &$({1},\hphantom{+}1)$
                        &$\hphantom{+}0$ &$-1$
                        &$-{1/2}$&$\hphantom{+}0$&$\hphantom{+}{1/2}$
                        &$(1, 1) $
                        &$+1$ &$\hphantom{+}0$ & $\hphantom{+}0$\\%
$                     $
                        &${\cal P}_{18}$                         
                        &$({1},\hphantom{+}1)$
                        &$\hphantom{+}0$ &$+1$
                        &$-{1/2}$&$\hphantom{+}0$&$\hphantom{+}{1/2}$
                        &$(1, 1) $
                        &$-1$ &$\hphantom{+}0$ & $\hphantom{+}0$\\%
\hline
$S+b_2+x+\alpha+e_5+e_6  $
                        &${\cal P}_{19}$                          
                        &$({1},\hphantom{+}1)$
                        &$\hphantom{+}0$ &$-1$
                        &$\hphantom{+}{1/2}$&$\hphantom{+}0$&$-{1/2}$
                        &$(1, 1) $
                        &$-1$ &$\hphantom{+}0$ & $\hphantom{+}0$\\%
$  $
                        &${\cal P}_{20}$               
                        &$({1},\hphantom{+}1)$
                        &$\hphantom{+}0$ &$+1$
                        &$\hphantom{+}{1/2}$&$\hphantom{+}0$&$-{1/2}$
                        &$(1, 1) $
                        &$+1$ &$\hphantom{+}0$ & $\hphantom{+}0$\\%
\hline
$S+b_2+x+\alpha+e_5  $
                        &${\cal P}_{21}$                 
                        &$({1},\hphantom{+}1)$
                        &$\hphantom{+}0$ &$-1$
                        &$\hphantom{+}{1/2}$&$\hphantom{+}0$&$-{1/2}$
                        &$(1, 1) $
                        &$+1$ &$\hphantom{+}0$ & $\hphantom{+}0$\\%
$  $
                        &${\cal P}_{22}$                  
                        &$({1},\hphantom{+}1)$
                        &$\hphantom{+}0$ &$+1$
                        &$\hphantom{+}{1/2}$&$\hphantom{+}0$&$-{1/2}$
                        &$(1, 1) $
                        &$-1$ &$\hphantom{+}0$ & $\hphantom{+}0$\\%
\hline
\end{tabular}
}
\caption{\label{tablePSe}\it
Twisted $SO(6)\times SO(4)$ exotic matter states and their charges under
the observable 
$SU(3)_C\times{SU(2)}_L\times{U(1)}_C\times U(1)_L\times{U(1)}^3$ 
and the hidden
${SU(4)}\times{SO(5)}\times U(1)^3$ 
gauge symmetries. These exotic states carry fractional electric charge
$\pm1/2$. 
}
\end{table}
    \end{landscape}
    \clearpage
}

\newpage
\normalsize
\afterpage{%
    \clearpage
    \begin{landscape}
        \centering 
\begin{table}
\noindent
{\small
\openup\jot
\begin{tabular}{|l|l|c|c|c|c|
                              c|c|c|
                                     c|c|c|c|}
\hline
sector&field&$SU(3)\times{SU(2)}$ &${U(1)}_C$ &$U(1)_L$
                                  &${U(1)}_1$&${U(1)}_2$&${U(1)}_3$
                                  &$SU(4)\times SO(5)$ 
                                  & $U(1)_4$& $U(1)_{5^\prime}$& $U(1)_{6^\prime}$\\
\hline
$S+b_1\pm\beta+z_1+(z_2)+e_3$
                         &${\cal F}_{1}$                  
                         &$({\bar3},\hphantom{+}1)$
                         &$\hphantom{+}{1}/{4}$ &$-{{1/2}}$
                         &$-{1/4}$&$\hphantom{+}{1/4}$&$\hphantom{+}{1/4}$
                         &$(1, 1) $
                         &$-{1/2}$ &$ -1$ & $-{1/2}$\\
$                      $
                         &${\cal F}_{2}$                    
                         &$({1},\hphantom{+}1)$
                         &$\hphantom{+}{3/4}$ &$\hphantom{+}{1/2}$
                         &$-{3/4}$&$-{1/4}$&$-{1/4}$
                         &$(1, 1) $
                         &$\hphantom{+}{1/2}$ &$ +1$ & $\hphantom{+}{1/2}$\\
\hline
$S+b_1\pm\beta+z_1+e_3+e_4$
                         &${\cal F}_{3}$                     
                         &$({\bar 3},\hphantom{+}1)$
                         &$\hphantom{+}{1}/{4}$ &$-{{1/2}}$
                         &$-{1/4}$&$\hphantom{+}{1/4}$&$\hphantom{+}{1/4}$
                         &$(1, 1) $
                      &$-{1/2}$&$+1$&$-{1/2}$\\
$ $
                         &${\cal F}_{4}$                     
                         &$({1},\hphantom{+}1)$
                         &$\hphantom{+}{3/4}$ &$\hphantom{+}{1/2}$
                         &$-{3/4}$&$-{1/4}$&$-{1/4}$
                         &$(1, 1) $
                      &$\hphantom{+}{1/2}$&$-1$&$\hphantom{+}{1/2}$\\
\hline
$S+b_1+x\pm\beta+z_1+e_5+e_6     $
                        &${\cal F}_{5}$                    
                        &$({3},\hphantom{+}1)$
                        &$-{1/4}$ &$\hphantom{+}{1/2}$
                        &$\hphantom{+}{1/4}$&$-{1/4}$&$-{1/4}$
                        &$(1, 1) $
                        &$-{1/2}$ &$+1$ & $-{1/2}$\\%
$     $
                        &${\cal F}_{6}$                        
                        &$({1},\hphantom{+}1)$
                        &$-{3/4}$ &$-{1/2}$
                  &$\hphantom{+}{3/4}$&$\hphantom{+}{1/4}$&$\hphantom{+}{1/4}$
                        &$(1, 1) $
                        &$\hphantom{+}{1/2}$ &$-1$ & $\hphantom{+}{1/2}$\\%
\hline
$S+b_1+x\pm\beta+z_1+z_2+e_4+e_5+e_6    $
                        &${\cal F}_{7}$                       
                        &$({3},\hphantom{+}1)$
                        &$-{1/4}$ &$\hphantom{+}{1/2}$
                        &$\hphantom{+}{1/4}$&$-{1/4}$&$-{1/4}$
                        &$(1, 1) $
                        &$-{1/2}$ &$-1$ & $-{1/2}$\\%
$                     $
                        &${\cal F}_{8}$                         
                        &$({1},\hphantom{+}1)$
                        &$-{3/4}$ &$-{1/2}$
                   &$\hphantom{+}{3/4}$&$\hphantom{+}{1/4}$&$\hphantom{+}{1/4}$
                        &$(1, 1) $
                        &$\hphantom{+}{1/2}$ &$+1$ & $\hphantom{+}{1/2}$\\%
\hline
$S+b_1\pm\beta+z_1+z_2+e_3+e_5$
                         &${\cal F}_{9}$                        
                         &$({1},\hphantom{+}2)$
                         &$-{3/4}$ &$\hphantom{+}{{1/2}}$
                         &$-{1/4}$&$\hphantom{+}{1/4}$&$\hphantom{+}{1/4}$
                         &$(1, 1) $
                      &$-{1/2}$&$-1$&$-{1/2}$\\
\hline
$S+b_1\pm\beta+z_1+e_3+e_4+e_5$
                         &${\cal F}_{10}$                  
                         &$({1},\hphantom{+}2)$
                         &$-{3/4}$ &$\hphantom{+}{{1/2}}$
                         &$-{1/4}$&$\hphantom{+}{1/4}$&$\hphantom{+}{1/4}$
                         &$(1, 1) $
                      &$-{1/2}$&$+1$&$-{1/2}$\\
\hline
$S+b_1+x\pm\beta+z_1+z_2+e_4+e_6    $
                        &${\cal F}_{11}$                      
                        &$({1},\hphantom{+}2)$
                        &$\hphantom{+}{3/4}$ &$-{1/2}$
                        &$\hphantom{+}{1/4}$&$-{1/4}$&$-{1/4}$
                        &$(1, 1) $
                        &$-{1/2}$ &$-1$ & $-{1/2}$\\%
\hline
$S+b_1+x\pm\beta+z_1+e_6     $
                        &${\cal F}_{12}$                      
                        &$({1},\hphantom{+}2)$
                        &$\hphantom{+}{3/4}$ &$-{1/2}$
                &$\hphantom{+}{1/4}$&$-{1/4}$&$-{1/4}$
                        &$(1, 1) $
                        &$-{1/2}$ &$+1$ & $-{1/2}$\\%
\hline
$S+b_3\pm\beta+(z_2)+e_6$
                         &${\cal F}_{13}$                       
                         &$({1},\hphantom{+}1)$
                         &$\hphantom{+}{3/4}$ &$\hphantom{+}{{1/2}}$
                         &$-{1/4}$&$\hphantom{+}{1/4}$&$-{1/4}$
                         &$(4, 1) $
                         &$-{1/2}$ &$ \hphantom{+}0$ & $-{1/2}$\\
\hline
$S+b_1\pm\beta+(z_2)+e_3+e_6$
                         &${\cal F}_{14}$                    
                         &$({1},\hphantom{+}1)$
                         &$-{3/4}$ &$-{{1/2}}$
                         &$-{1/4}$&$\hphantom{+}{1/4}$&$\hphantom{+}{1/4}$
                         &$(4, 1) $
                      &$\hphantom{+}{1/2}$&$\hphantom{+}0$&$\hphantom{+}{1/2}$\\
\hline
$S+b_1\pm\beta+(z_2)+e_3+e_4$
                         &${\cal F}_{15}$                        
                         &$({1},\hphantom{+}1)$
                         &$\hphantom{+}{3/4}$ &$\hphantom{+}{1/2}$
                         &$\hphantom{+}{1/4}$&$-{1/4}$&$\hphantom{+}{1/4}$
                         &$(4, 1) $
                      &$-{1/2}$&$\hphantom{+}0$&$-{1/2}$\\
\hline
$S+b_2\pm\beta+(z_2)+e_1+e_5+e_6$
                         &${\cal F}_{16}$                    
                         &$({1},\hphantom{+}1)$
                         &$\hphantom{+}{3/4}$ &$\hphantom{+}{1/2}$
                         &$-{1/4}$&$\hphantom{+}{1/4}$&$-{1/4}$
                         &$({4}, 1) $
                         &$-{1/2}$ &$ \hphantom{+}0$ & $-{1/2}$\\
\hline
$S+b_2+x\pm\beta+(z_2)+e_1$
                         &${\cal F}_{17}$                       
                         &$({1},\hphantom{+}1)$
                         &$-{3/4}$ &$-{1/2}$
                         &$\hphantom{+}{1/4}$&$-{1/4}$&$\hphantom{+}{1/4}$
                         &$({\bar4}, 1) $
                         &$-{1/2}$ &$ \hphantom{+}0$ & $-{1/2}$\\
\hline
$S+b_1+x\pm\beta+(z_2)+e_4+e_5+e_6     $
                        &${\cal F}_{18}$                       
                        &$({1},\hphantom{+}1)$
                        &$-{3/4}$ &$-{1/2}$
                        &$-{1/4}$&$\hphantom{+}{1/4}$&$\hphantom{+}{1/4}$
                        &$({\bar4}, 1) $
                        &$-{1/2}$ &$\hphantom{+}0$ & $-{1/2}$\\%
\hline
$S+b_2+x\pm\beta+(z_2)+e_5     $
                        &${\cal F}_{19}$                    
                        &$({1},\hphantom{+}1)$
                        &$-{3/4}$ &$-{1/2}$
                        &$\hphantom{+}{1/4}$&$-{1/4}$&$\hphantom{+}{1/4}$
                        &$({\bar4}, 1) $
                        &$-{1/2}$ &$\hphantom{+}0$ & $-{1/2}$\\%
\hline
$S+b_1+x\pm\beta+(z_2)+e_5     $
                        &${\cal F}_{20}$                       
                        &$({1},\hphantom{+}1)$
                        &$\hphantom{+}{3/4}$ &$\hphantom{+}{1/2}$
                &$\hphantom{+}{1/4}$&$-{1/4}$&$-{1/4}$
                        &$({\bar4}, 1) $
                &$\hphantom{+}{1/2}$&$\hphantom{+}0$&$\hphantom{+}{1/2}$\\%
\hline
$S+b_1\pm\beta+z_1+z_2+e_3+e_6$
                         &${\cal F}_{21}$                     
                         &$({1},\hphantom{+}1)$
                         &$-{3/4}$ &$-{{1/2}}$
                         &$-{1/4}$&$\hphantom{+}{1/4}$&$-{3/4}$
                         &$(1, 1) $
                      &$-{1/2}$&$-1$&$-{1/2}$\\
$ $
                         &${\cal F}_{22}$                   
                         &$({1},\hphantom{+}1)$
                         &$\hphantom{+}{3/4}$ &$\hphantom{+}{{1/2}}$
                         &$\hphantom{+}{1/4}$&$\hphantom{+}{3/4}$&$-{1/4}$
                         &$(1, 1) $
                      &$\hphantom{+}{1/2}$&$+1$&$\hphantom{+}{1/2}$\\
\hline
$S+b_1\pm\beta+z_1+e_3+e_4+e_6$
                         &${\cal F}_{23}$                    
                         &$({1},\hphantom{+}1)$
                         &$-{3/4}$ &$-{{1/2}}$
                         &$-{1/4}$&$\hphantom{+}{1/4}$&$-{3/4}$
                         &$(1, 1) $
                      &$-{1/2}$&$+1$&$-{1/2}$\\
$ $
                         &${\cal F}_{24}$                  
                         &$({1},\hphantom{+}1)$
                         &$\hphantom{+}{3/4}$ &$\hphantom{+}{1/2}$
                         &$\hphantom{+}{1/4}$&$\hphantom{+}{3/4}$&$-{1/4}$
                         &$(1, 1) $
                      &$\hphantom{+}{1/2}$&$-1$&$\hphantom{+}{1/2}$\\
\hline
$S+b_1+x\pm\beta+z_1+z_2+e_4+e_5     $
                        &${\cal F}_{25}$                       
                        &$({1},\hphantom{+}1)$
                        &$\hphantom{+}{3/4}$ &$\hphantom{+}{1/2}$
                        &$\hphantom{+}{1/4}$&$-{1/4}$&$\hphantom{+}{3/4}$
                        &$(1, 1) $
                        &$-{1/2}$ &$-1$ & $-{1/2}$\\%
$     $
                        &${\cal F}_{26}$                  
                        &$({1},\hphantom{+}1)$
                        &$-{3/4}$ &$-{1/2}$
                        &$-{1/4}$&$-{3/4}$&$\hphantom{+}{1/4}$
                        &$({1}, 1) $
                        &$\hphantom{+}{1/2}$ &$+1$ & $\hphantom{+}{1/2}$\\%
\hline
$S+b_1+x\pm\beta+z_1+e_5     $
                        &${\cal F}_{27}$                    
                        &$({1},\hphantom{+}1)$
                        &$\hphantom{+}{3/4}$ &$\hphantom{+}{1/2}$
                &$\hphantom{+}{1/4}$&$-{1/4}$&$\hphantom{+}{3/4}$
                        &$(1, 1) $
                        &$-{1/2}$ &$+1$ & $-{1/2}$\\%
$     $
                        &${\cal F}_{28}$                        
                        &$({1},\hphantom{+}1)$
                        &$-{3/4}$ &$-{1/2}$
                  &$-{1/4}$&$-{3/4}$&$\hphantom{+}{1/4}$
                        &$(1, 1) $
                        &$\hphantom{+}{1/2}$ &$-1$ & $\hphantom{+}{1/2}$\\%
\hline
\end{tabular}
}
\caption{\label{tablefsu5e}\it
Twisted $SU(5)\times U(1)$ exotic matter states and their charges under
the observable 
$SU(3)_C\times{SU(2)}_L\times{U(1)}_C\times U(1)_L\times{U(1)}^3$ 
and the hidden
${SU(4)}\times{SO(5)}\times U(1)^3$ 
gauge symmetries. These exotic states carry fractional electric charge
$\pm1/2$.
}
\end{table}
    \end{landscape}
    \clearpage
}

\newpage
\normalsize
\afterpage{%
    \clearpage
    \begin{landscape}
        \centering 
\begin{table}
\noindent
{\small
\openup\jot
\begin{tabular}{|l|l|c|c|c|c|
                              c|c|c|
                                     c|c|c|c|}
\hline
sector&field&$SU(3)\times{SU(2)}$ &${U(1)}_C$ &$U(1)_L$
                                     &${U(1)}_1$&${U(1)}_2$&${U(1)}_3$
                                     &$SU(4)\times SO(5)$ 
                                     & $U(1)_4$& $U(1)_{5^\prime}$& $U(1)_{6^\prime}$\\
\hline
$S+b_3+x+(z_2)$
                        &${V}_{1}$
                        &$({1},\hphantom{+}1)$
                        &$\hphantom{+}0$ &$\hphantom{+}0$
                        &$-{1/2}$&$\hphantom{+}{1/2}$&$\hphantom{+}0$
                        &$(6, 1) $
                        &$\hphantom{+}0$ &$\hphantom{+}0$ & $\hphantom{+}0$\\%
\hline
$S+b_3+x+(z_2)+e_3       $
                        &${V}_{2}$               
                        &$({1},\hphantom{+}1)$
                        &$\hphantom{+}0$ &$\hphantom{+}0$
                        &$\hphantom{+}{1/2}$&$-{1/2}$&$\hphantom{+}0$
                        &$(6, 1) $
                        &$\hphantom{+}0$ &$\hphantom{+}0$ & $\hphantom{+}0$\\%
\hline
$S+b_2+x+(z_2)+e_1+e_6  $
                        &${V}_{3}$              
                        &$({1},\hphantom{+}1)$
                        &$\hphantom{+}0$ &$\hphantom{+}0$
                        &$-{1/2}$&$\hphantom{+}0$&$\hphantom{+}{1/2}$
                        &$(6, 1) $
                        &$\hphantom{+}0$ &$\hphantom{+}0$ & $\hphantom{+}0$\\%
\hline
$S+b_1+x+(z_2)+e_4+e_5  $
                        &${V}_{4}$              
                        &$({1},\hphantom{+}1)$
                        &$\hphantom{+}0$ &$\hphantom{+}0$
                        &$\hphantom{+}0$&$-{1/2}$&$\hphantom{+}{1/2}$
                        &$(6, 1) $
                        &$\hphantom{+}0$ &$\hphantom{+}0$ & $\hphantom{+}0$\\%
\hline
$S+b_2+x+(z_2)+e_5+e_6  $
                        &${V}_{5}$             
                        &$({1},\hphantom{+}1)$
                        &$\hphantom{+}0$ &$\hphantom{+}0$
                        &$-{1/2}$&$\hphantom{+}0$&$\hphantom{+}{1/2}$
                        &$(6, 1) $
                        &$\hphantom{+}0$ &$\hphantom{+}0$ & $\hphantom{+}0$\\%
\hline
$S+b_1+2\beta+z_1+(z_2) +e_3$
                        &${V}_{6}$               
                        &$({1},\hphantom{+}1)$
                        &$\hphantom{+}0$ &$\hphantom{+}0$
                        &$\hphantom{+}0$&$-{1/2}$&$-{1/2}$
                        &$({\bar4}, 1 ) $
                        &$\hphantom{+}0$ &$-1$ & $\hphantom{+}0$\\
\hline
$S+b_1+2\beta+z_1+(z_2)+e_3+e_4$
                         &${V}_{7}$              
                         &$({1},\hphantom{+}1)$
                         &$\hphantom{+}0$ &$\hphantom{+}0$
                         &$\hphantom{+}0$&$-{1/2}$&$-{1/2}$
                         &$({\bar4}, 1 ) $
                      &$\hphantom{+}0$&$+1$&$\hphantom{+}0$\\
\hline
$S+b_3+2\beta+z_1+(z_2)+e_1+e_4$    
                        &${V}_{8}$             
                        &$({1},\hphantom{+}1)$
                        &$\hphantom{+}0$ &$\hphantom{+}0$
                        &$-{1/2}$&$-{1/2}$&$\hphantom{+}0$
                        &$({\bar4}, 1) $
                        &$\hphantom{+}0$ &$-1$ & $\hphantom{+}0$\\%
\hline
$S+b_3+2\beta+z_1+(z_2)+e_1$
                        &${V}_{9}$       
                        &$({1},\hphantom{+}1)$
                        &$\hphantom{+}0$ &$\hphantom{+}0$
                        &$-{1/2}$&$-{1/2}$&$\hphantom{+}0$
                        &$({\bar4}, 1) $
                        &$\hphantom{+}0$ &$+1$ & $\hphantom{+}0$\\%
\hline
$S+b_1+2\beta+z_1+(z_2)+e_3+e_6$
                        &${V}_{10}$             
                        &$({1},\hphantom{+}1)$
                        &$\hphantom{+}0$ &$\hphantom{+}0$
                        &$\hphantom{+}0$&$-{1/2}$&$-{1/2}$
                        &$(4, 1) $
                        &$\hphantom{+}0$ &$+1$ & $\hphantom{+}0$\\
\hline
$S+b_1+2\beta+z_1+(z_2)+e_3+e_4+e_6$
                         &${V}_{11}$         
                         &$({1},\hphantom{+}1)$
                         &$\hphantom{+}0$ &$\hphantom{+}0$
                         &$\hphantom{+}0$&$-{1/2}$&$-{1/2}$
                         &$({4}, 1 ) $
                      &$\hphantom{+}0$&$-1$&$\hphantom{+}0$\\
\hline
$S+b_3+2\beta+z_1+(z_2)+e_1+e_3$    
                        &${V}_{12}$             
                        &$({1},\hphantom{+}1)$
                        &$\hphantom{+}0$ &$\hphantom{+}0$
                      &$\hphantom{+}{1/2}$&$\hphantom{+}{1/2}$&$\hphantom{+}0$
                        &$(4, 1) $
                        &$\hphantom{+}0$ &$-1$ & $\hphantom{+}0$\\%
\hline
$S+b_3+2\beta+z_1+(z_2)+e_1+e_3+e_4$    
                        &${V}_{13}$             
                        &$({1},\hphantom{+}1)$
                        &$\hphantom{+}0$ &$\hphantom{+}0$
                      &$\hphantom{+}{1/2}$&$\hphantom{+}{1/2}$&$\hphantom{+}0$
                        &$(4, 1) $
                        &$\hphantom{+}0$ &$+1$ & $\hphantom{+}0$\\%
\hline
$S+b_3+x+(z_3)+e_1+e_2$        
                        &${V}_{14}$              
                        &$({1},\hphantom{+}1)$
                        &$\hphantom{+}0$ &$\hphantom{+}0$
                        &$-{1/2}$&$-{1/2}$&$\hphantom{+}0$
                        &$(1, 4) $
                        &$-1$ &$\hphantom{+}0$ & $\hphantom{+}0$\\%
\hline
$S+b_3+x+(z_3)+e_1+e_2+e_3$    
                        &${V}_{15}$             
                        &$({1},\hphantom{+}1)$
                        &$\hphantom{+}0$ &$\hphantom{+}0$
                        &$-{1/2}$&$-{1/2}$&$\hphantom{+}0$
                        &$(1, 4) $
                        &$-1$ &$\hphantom{+}0$ & $\hphantom{+}0$\\%
\hline
$S+b_3+z_2+(z_3)+e_2+e_3$        
                        &${V}_{16}$             
                        &$({1},\hphantom{+}1)$
                        &$\hphantom{+}0$ &$\hphantom{+}0$
                        &$-{1/2}$&$-{1/2}$&$\hphantom{+}0$
                        &$(1, 4) $
                        &$\hphantom{+}0$ &$\hphantom{+}0$ & $-1$\\%
\hline
$S+b_3+z_2+(z_3)+e_2$    
                        &${V}_{17}$               
                        &$({1},\hphantom{+}1)$
                        &$\hphantom{+}0$ &$\hphantom{+}0$
                        &$-{1/2}$&$-{1/2}$&$\hphantom{+}0$
                        &$(1, 4) $
                        &$\hphantom{+}0$ &$\hphantom{+}0$ & $-1$\\%
\hline
$S+b_1+2\beta+(z_3)+e_3+e_5+e_6$
                         &${V}_{18}$              
                         &$({1},\hphantom{+}1)$
                         &$\hphantom{+}0$ &$\hphantom{+}0$
                         &$\hphantom{+}0$&$-{1/2}$&$\hphantom{+}{1/2}$
                         &$(1, 5) $
                      &$\hphantom{+}0$&$\hphantom{+}0$&$\hphantom{+}0$\\
\hline
$S+b_1+2\beta+z_2+e_3$
                        &${V}_{19}$               
                        &$({1},\hphantom{+}1)$
                        &$\hphantom{+}0$ &$\hphantom{+}0$
                        &$\hphantom{+}0$&$-{1/2}$&$\hphantom{+}{1/2}$
                        &$(1, 1) $
                        &$-1$ &$\hphantom{+}0$ & $-1$\\
$                      $
                        &${V}_{20}$               
                        &$({1},\hphantom{+}1)$
                        &$\hphantom{+}0$ &$\hphantom{+}0$
                        &$\hphantom{+}0$&$-{1/2}$&$\hphantom{+}{1/2}$
                        &$(1, 1) $
                        &$+1$ &$\hphantom{+}0$ & $+1$\\
\hline
$S+b_1+2\beta+z_2+e_3+e_6$
                        &${V}_{21}$               
                        &$({1},\hphantom{+}1)$
                        &$\hphantom{+}0$ &$\hphantom{+}0$
                        &$\hphantom{+}0$&$-{1/2}$&$\hphantom{+}{1/2}$
                        &$(1, 1) $
                        &$-1$ &$\hphantom{+}0$ & $+1$\\
$                      $
                        &${V}_{22}$               
                        &$({1},\hphantom{+}1)$
                        &$\hphantom{+}0$ &$\hphantom{+}0$
                        &$\hphantom{+}0$&$-{1/2}$&$\hphantom{+}{1/2}$
                        &$(1, 1) $
                        &$+1$ &$\hphantom{+}0$ & $-1$\\
\hline
$S+b_3+2\beta+z_2+e_1+e_3+e_4$    
                        &${V}_{23}$             
                        &$({1},\hphantom{+}1)$
                        &$\hphantom{+}0$ &$\hphantom{+}0$
                        &$-{1/2}$&$\hphantom{+}{1/2}$&$\hphantom{+}0$
                        &$(1, 1) $
                        &$-1$ &$\hphantom{+}0$ & $-1$\\%
$ $    
                        &${V}_{24}$              
                        &$({1},\hphantom{+}1)$
                        &$\hphantom{+}0$ &$\hphantom{+}0$
                        &$-{1/2}$&$\hphantom{+}{1/2}$&$\hphantom{+}0$
                        &$(1, 1) $
                        &$+1$ &$\hphantom{+}0$ & $+1$\\%
\hline
$S+b_3+2\beta+z_2+e_1+e_4$    
                        &${V}_{25}$             
                        &$({1},\hphantom{+}1)$
                        &$\hphantom{+}0$ &$\hphantom{+}0$
                        &$\hphantom{+}{1/2}$&$-{1/2}$&$\hphantom{+}0$
                        &$(1, 1) $
                        &$-1$ &$\hphantom{+}0$ & $-1$\\%
$ $    
                        &${V}_{26}$             
                        &$({1},\hphantom{+}1)$
                        &$\hphantom{+}0$ &$\hphantom{+}0$
                        &$\hphantom{+}{1/2}$&$-{1/2}$&$\hphantom{+}0$
                        &$(1, 1) $
                        &$+1$ &$\hphantom{+}0$ & $+1$\\%
\hline
$S+b_2+x+z_2+e_1      $
                        &${V}_{27}$               
                        &$({1},\hphantom{+}1)$
                        &$\hphantom{+}0$ &$\hphantom{+}0$
                        &$-{1/2}$&$\hphantom{+}0$&$\hphantom{+}{1/2}$
                        &$(1, 1) $
                        &$\hphantom{+}0$ &$-2$ & $\hphantom{+}0$\\%
$                     $
                        &${V}_{28}$               
                        &$({1},\hphantom{+}1)$
                        &$\hphantom{+}0$ &$\hphantom{+}0$
                        &$-{1/2}$&$\hphantom{+}0$&$\hphantom{+}{1/2}$
                        &$(1, 1) $
                        &$\hphantom{+}0$ &$+2$ & $\hphantom{+}0$\\%
\hline
$S+b_1+x+z_2+e_4+e_5+e_6$    
                        &${V}_{29}$               
                        &$({1},\hphantom{+}1)$
                        &$\hphantom{+}0$ &$\hphantom{+}0$
                        &$\hphantom{+}0$&$-{1/2}$&$\hphantom{+}{1/2}$
                        &$(1, 1) $
                        &$\hphantom{+}0$ &$-2$ & $\hphantom{+}0$\\%
$ $    
                        &${V}_{30}$               
                        &$({1},\hphantom{+}1)$
                        &$\hphantom{+}0$ &$\hphantom{+}0$
                        &$\hphantom{+}0$&$-{1/2}$&$\hphantom{+}{1/2}$
                        &$(1, 1) $
                        &$\hphantom{+}0$ &$+2$ & $\hphantom{+}0$\\%
\hline
$S+b_2+x+z_2+e_5$    
                        &${V}_{31}$                 
                        &$({1},\hphantom{+}1)$
                        &$\hphantom{+}0$ &$\hphantom{+}0$
                        &$-{1/2}$&$\hphantom{+}0$&$\hphantom{+}{1/2}$
                        &$(1, 1) $
                        &$\hphantom{+}0$ &$-2$ & $\hphantom{+}0$\\%
$ $    
                        &${V}_{32}$               
                        &$({1},\hphantom{+}1)$
                        &$\hphantom{+}0$ &$\hphantom{+}0$
                        &$-{1/2}$&$\hphantom{+}0$&$\hphantom{+}{1/2}$
                        &$(1, 1) $
                        &$\hphantom{+}0$ &$+2$ & $\hphantom{+}0$\\%
\hline
$S+(z_3)$    
                        &${V}_{33}$                 
                        &$({1},\hphantom{+}1)$
                        &$\hphantom{+}0$ &$\hphantom{+}0$
                        &$\hphantom{+}0$&$\hphantom{+}0$&$\hphantom{+}0$
                        &$(1, 5) $
                        &$\hphantom{+}0$ &$\hphantom{+}0$ & $\hphantom{+}0$\\%
\hline
\end{tabular}
}
\caption{\label{tableso10singlets}\it
Twisted $SO(10)$ singlet matter states and their charges under
the observable 
$SU(3)_C\times{SU(2)}_L\times{U(1)}_C\times U(1)_L\times{U(1)}^3$ 
and the hidden
${SU(4)}\times{SO(5)}\times U(1)^3$ 
gauge symmetries.
The last state, $V_{33}$, is an untwisted matter 
state charged under the hidden sector.
}
\end{table}
    \end{landscape}
    \clearpage
}

\section{\emph{Conclusions}}\label{conc}

In this paper we extended the free fermionic classifications methodology to 
the class of Standard--like Models, in which the $SO(10)$ GUT symmetry is 
broken at the string level to $SU(3)\times SU(2)\times U(1)^2$. The 
SLM free fermionic heterotic--string models uniquely require 
two basis vectors that break the $SO(10)$ symmetry. One to the PS subgroup 
and one to the FSU5 subgroup. Compared to the PS and FSU5 cases this 
substantially increases the computational complexity of the analysis
of the SLM vacua. Compared to 
the classifications of the corresponding PS and FSU5 models, the space
of SLM models is vastly increased. To extract the phenomenologically 
interesting three generation models therefore required adaptation of the 
methodology. Rather than generating random phases of the entire space 
of vacua, we divided the process in two steps. The first involves imposing 
a set of constraints on the GGSO phases of models with $SO(10)$ GUT symmetry. 
These constraints pre--select $SO(10)$ GGSO configurations that yield 
three generation models, and are dubbed as fertile $SO(10)$ models. At this
level we generate random choices of GGSO configurations that satisfy the 
fertility constraints. To these preselected configurations we add 
the $SO(10)$ breaking basis vectors, and perform a complete classification
of the additional GGSO coefficients, resulting in some $10^7$ three 
generation Standard--like Models. Additionally, we discussed the 
imposition of a viable top quark Yukawa couplings in the selection process, 
and the reduction of model degeneracy, which is obtained by fixing the 
$Z_2\times Z_2$ orbifold planes from which the top quark left and right
components are obtained. We further used our computerised method to 
explore in detail an exemplary three generation SLM model with special 
phenomenological properties, which were not obtained in previous SLM
constructions. The first is the presence of an additional pair of 
right--handed neutrino and its conjugated field. These can be used as 
heavy Higgs representations to break the additional $U(1)_{Z^\prime}$
of eq. (\ref{u1zprime}) along supersymmetric flat directions, whereas 
in earlier models exotic states with fractional $U(1)_{Z^\prime}$
charge were used for that purpose \cite{slm}. The consequence is that 
in our new model one can realise the dark matter scenario discussed 
in ref. \cite{ccr}, whereas this was not possible in the models
of refs. \cite{slm}. The second distinct property of our new model
is the existence of an additional pair of $Q$, ${\bar Q}$ states 
that may be instrumental in mitigating the heterotic--string 
gauge coupling unification problem \cite{gutu, gcu}. 

The methodology developed in this work therefore enables us to 
generate a larger number of phenomenologically viable SLM 
free fermionic heterotic--string vacua, as compared to the earlier
trial--and--error method of \cite{slm}. One can envision using this 
method to delve deeper in the phenomenological detail. In this 
paper we focused on the analysis of the observable Standard Model 
matter states. Analysis of the enhanced symmetries and 
exotic states can be further developed, along the lines of 
earlier classifications \cite{acfkr, frs, hasan, HasansThesis}.
Furthermore, the vast space of GGSO configurations entailed 
that our analysis here is slated toward models that can produce 
phenomenologically viable models. It would therefore be of interest
to develop alternative computerised methods, such as those developed in 
refs. \cite{alternativecm}, and to explore the symmetries underlying 
the larger space of vacua. 


\section*{Acknowledgments}
AEF thanks the theoretical physics departments at CERN, Oxford
University and Weizmann Institute for hospitality.
AEF is supported in part by the STFC (ST/L000431/1).

\end{document}